\PassOptionsToPackage{dvipsnames}{xcolor}
\documentclass[%
article, 
twocolumn,
superscriptaddress,
preprintnumbers,
nofootinbib,
amsmath,amssymb,
aps,
longbibliography]{revtex4-2}
\usepackage{graphicx}
\usepackage{dcolumn}
\usepackage{bm}
\usepackage{bbold}
\usepackage{calc}
\usepackage{algpseudocode}
\usepackage{changes}
\usepackage[braket, qm]{qcircuit}
\usepackage{xcolor}
\usepackage{braket}
\usepackage{amsthm}
\usepackage{ulem}
\usepackage{url}
\usepackage{hyperref}

\graphicspath{{./figures/}}
\usepackage{url}

\usepackage{colortbl}
\usepackage{comment}

\newcolumntype{L}[1]{>{\raggedright\arraybackslash}p{#1}}

\newcommand {\nc} {\newcommand}
\nc {\SY} [1]{\textcolor{magenta}{#1}}
\nc {\AB} [1]{\textcolor{blue}{#1}}
\nc {\ER} [1]{\textcolor{orange}{#1}}
\nc {\TM} [1]{\textcolor{ForestGreen}{#1}} 

\hypersetup{
    colorlinks=true,
    linkcolor=blue,
    citecolor=blue,
    filecolor=magenta,      
    urlcolor=blue,
    pdftitle={Nuclear Many-Body Systems as Benchmarks for Quantum Computing},
    pdfpagemode=FullScreen,
    }

\begin{document}

\preprint{RIKEN-iTHEMS-Report-26}

\title{Nuclear Many-Body Systems as Benchmarks for Quantum Computing}

\author{Sota Yoshida}
  \affiliation{School of Data Science and Management, Utsunomiya University, Mine, Utsunomiya, 321-8505, Japan}
  \affiliation{RIKEN Nishina Center for Accelerator-based Science, RIKEN, Wako 351-0198, Japan}
\author{Alessandro Baroni}
  \affiliation{National Center for Computational Sciences, Oak Ridge National Laboratory, TN 37831, USA}
\author{Takayuki Miyagi}
  \affiliation{Center for Computational Sciences, University of Tsukuba, Japan}
\author{Ermal Rrapaj}
  \affiliation{National Energy Research Scientific Computing Center, Lawrence Berkeley National Laboratory, Berkeley, CA 94720, USA }
  \affiliation{Department of Physics, University of California, Berkeley, CA 94720, USA}
  \affiliation{RIKEN iTHEMS, Wako, Saitama 351-0198, Japan}

\begin{abstract}
We present NuQuLib, an open-source framework for benchmarking quantum algorithms for nuclear many-body systems. NuQuLib uses the standard Jordan–Wigner transformation to connect second-quantized nuclear Hamiltonians with quantum algorithm implementations and resource estimates. The examples span valence and no-core shell-model spaces and include two-nucleon interactions and selected cases with three-nucleon forces.
We estimate how logical T-gate counts grow with model-space size for quantum phase estimation (QPE), quantum Krylov subspace methods, and observable dynamic mode decomposition at fixed algorithm parameters.
In a 40-qubit no-core space, three-nucleon forces increase the T-gate count of one first-order Trotter step by a factor of about 270.
For QPE, we also compare two implementations at a common absolute energy accuracy of 100 keV relative to each Hamiltonian, using pairwise commutator coefficients evaluated directly for spaces of up to 880 qubits.
Across the 20 Hamiltonians evaluated, qubitization gives lower estimated T-gate counts per spectral sample than
first-order Trotterization with step counts set by the pairwise commutator bound.
Finite-shot simulations of ground-state energy estimation in ${}^{6}$He and ${}^{22}$O illustrate
a trade-off between sampling and coherent evolution. For the initial states and
measurement protocols tested, QPE used the fewest circuit executions per estimate
and the most sequential Trotter substeps in a single circuit.
\end{abstract}

\maketitle

\section{Introduction}
\label{sec:intro}

Quantum hardware technology has seen significant developments in the last decade~\cite{doi:10.1126/science.adz8659, SETR2025}, in conjunction with algorithms for quantum simulation and related tasks~\cite{cao2019,bauer2020quantum,Gilyen19,fable,PRXQuantum.5.037001}.
For such progress to translate to applications in the physical sciences, however, one also needs domain-specific formulations that map realistic problems into qubit Hamiltonians and benchmark suites that make different algorithmic choices comparable on common ground.
Quantum chemistry has become a leading example of this development, thanks to both the scientific relevance of molecular simulation and the maturity of its benchmark culture for ground-state energy estimation~\cite{reiher2017,Goings:2022}.
That benchmark-driven perspective has already led to substantial algorithmic refinements and reduced quantum resource estimates~\cite{Rocca:2024}.

Nuclear Hamiltonians offer a complementary setting for benchmarking quantum algorithms.
Atomic nuclei are self-bound, strongly correlated many-body systems whose
interactions depend on spin and isospin and respect parity and time-reversal symmetry.
Chiral effective field theory (ChEFT) organizes two- and three-nucleon forces
in a systematic expansion~\cite{MACHLEIDT:2024}.
Previous work includes proof-of-principle variational calculations for light
and medium-mass nuclei~\cite{PhysRevLett.120.210501,PhysRevC.106.034325,Perez:2023,PhysRevC.108.064305,PhysRevC.105.064308}.
Quantum simulation algorithms and resource estimates have also been developed
for nuclear effective field theories on spatial lattices~\cite{Watson2026,Spagnoli2026}.
Other studies have addressed scattering and reactions~\cite{Wang:2024,Turro:2024,Rethinasamy:2026},
resonances~\cite{Zhang:2025,Singh:2025}, and early fault-tolerant simulation
of nuclear systems~\cite{Gu:2026,Gibbs:2026,Du:2026,Sakuma:2026}.
Here we use shell-model Hamiltonians in a harmonic-oscillator basis,
in valence spaces and in no-core spaces that include three-nucleon forces,
as common inputs for comparing several eigenvalue algorithms under a
single cost model.

We introduce NuQuLib~\cite{Repo_NuQuLib,*Zenodo_NuQuLib}, a framework
that connects nuclear-structure inputs to qubit Hamiltonians, quantum
circuits, and resource estimates for eigenvalue algorithms.
Three-nucleon forces add many terms to the encoded Hamiltonian,
which allows us to examine how they affect the cost of quantum time
evolution. The main elements of this work are as follows.
\begin{itemize}
\item A workflow that maps valence-space and no-core shell-model
Hamiltonians, including no-core cases with three-nucleon forces, to
Jordan--Wigner qubit Hamiltonians, together with the accompanying
data (Secs.~\ref{sec:workflow} and \ref{sec:Nuc_Hamiltonian}).
\item Structural Pauli-term counts and measurement-grouping benchmarks
for these Hamiltonians (Secs.~\ref{subsec:Encoding} and
\ref{sec:measurement_grouping}).
\item Evaluation of the first-order pairwise commutator coefficient
from all retained Hamiltonian terms by summing weights over shared
subsets of Majorana indices. This avoids explicit pair enumeration
and allows us to treat spaces of up to 880 qubits with NN interactions
and 140 qubits with NN+3N interactions
(Sec.~\ref{sec:pairwise_bounds} and Appendix~\ref{app:pair_coefficient}).
\item Logical T-gate estimates at fixed algorithm parameters for quantum
phase estimation (QPE)~\cite{Kitaev95}, quantum Krylov subspace methods
(QKrylov)~\cite{Cortes_PRA22}, and observable dynamic mode
decomposition (ODMD)~\cite{ODMD_Shen}, and a comparison of first-order
Trotterized and qubitized QPE at $\Delta E=100$~keV relative to each
Hamiltonian (Secs.~\ref{sec:Qresource_summary} and
\ref{sec:qpe_fixed_accuracy}).
\item Finite-shot calculations for ${}^{6}$He and ${}^{22}$O that
illustrate the trade-off between the total number of shots and the
longest coherent evolution used in a single circuit to estimate the
ground-state energy from a specified initial state
(Sec.~\ref{sec:finite_shot}).
\end{itemize}

The paper is organized as follows.
Section~\ref{sec:workflow} describes the NuQuLib workflow and the
benchmark tasks. Section~\ref{sec:Nuc_Hamiltonian} specifies the nuclear
Hamiltonians, model spaces, and qubit encoding, and
Sec.~\ref{sec:quantum_algorithms} summarizes QPE, QKrylov, and ODMD.
Section~\ref{sec:resource_estimation} defines the cost model, including
the pairwise commutator coefficient and measurement grouping.
Section~\ref{sec:results} presents the resource estimates: scaling at
fixed algorithm parameters, QPE at fixed energy accuracy, finite-shot
ground-state energy estimates, and the overhead of state preparation.
Section~\ref{sec:demonstrations} illustrates the workflow on systems
accessible to classical simulation, and Sec.~\ref{sec:outlook}
discusses limitations and future directions.
The appendices give the T-gate count formulas
(Appendix~\ref{app:Tformulas}), the Majorana-based evaluation of
commutator coefficients and measurement settings
(Appendix~\ref{app:pair_bound}), the qubitized QPE cost model
(Appendix~\ref{app:qubitization}), QSVT-based time evolution
(Appendix~\ref{app:qsvt_time_evolution}), details of the finite-shot
calculations (Appendix~\ref{app:r21_fixed_sampling}),
the quantum projection filter as a possible
state-preparation method (Appendix~\ref{app:qpf}),
a higher-order product-formula test (Appendix~\ref{app:higher_order}), and a
variational example (Appendix~\ref{app:vqe_demo}).

\begin{figure*}
\includegraphics[width=0.98\textwidth]{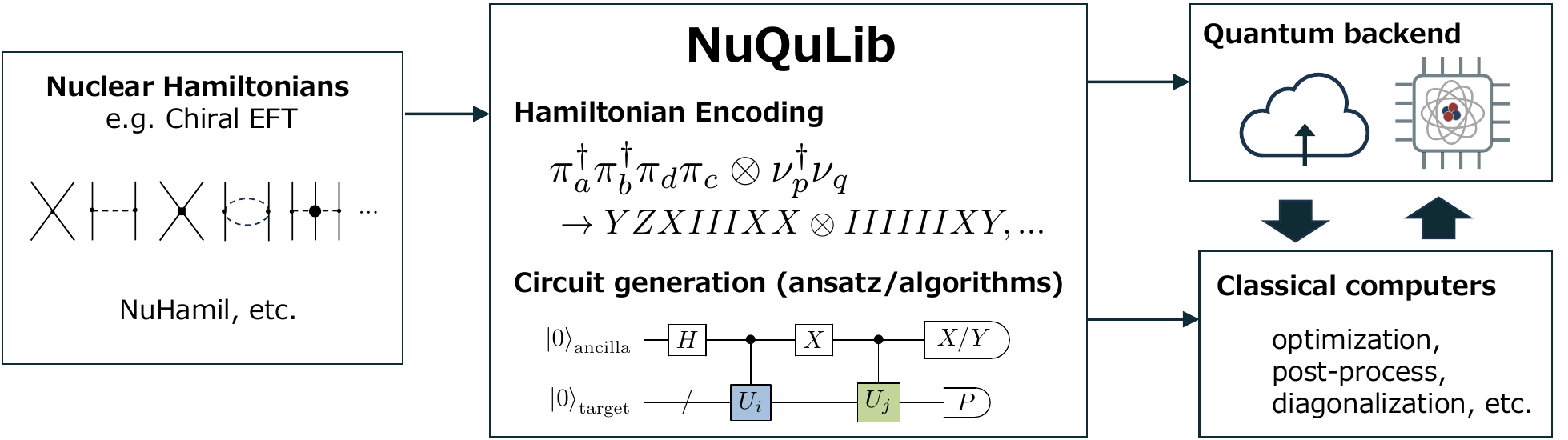}
\caption{NuQuLib workflow.
Starting from nuclear Hamiltonian matrix elements in a chosen model space,
NuQuLib constructs the qubit Hamiltonian and generates circuits for state preparation and
quantum algorithms. The right-hand side shows how circuit execution is
combined with classical processing, such as optimization or diagonalization.
\label{fig:Sketch_Workflow}
}
\end{figure*}

\begin{table*}
\caption{Benchmark instances used in this work.
\textit{Resource estimates:} each encoded Hamiltonian is analyzed as an
operator on the full qubit space, without preparing a nuclear state or
executing a circuit. $N_q$ counts system qubits, one per single-particle
state. In the no-core spaces, $e_{\max}$ bounds the oscillator quanta of
the single-particle states ($2n+l\le e_{\max}$) and $E_{3\max}$ those of
the three-body matrix elements; no many-body $N_{\max}$ truncation is
imposed.
\textit{Statevector simulations:} ``Compared with'' gives the reference
for each result: exact diagonalization in the same space and symmetry
sector (projection $M$, parity $\pi$), the doubly occupied configuration
interaction (DOCI) solution in the zero-seniority pair space, or the full
configuration interaction (FCI) energy. $P(J=0)$ is the weight of the
$J=0$ component. For the finite-shot estimates, 100~keV and 95\% are the
energy window and the target success probability.
The valence interactions are CKpot~\cite{CohenKurath}, USDB~\cite{USDB},
YSOX~\cite{Yuan2012}, GXPF1A~\cite{Honma2005}, SDPF-MU~\cite{Utsuno2012},
and SDPFSDG-MU~\cite{Utsuno2012PTP}. The no-core Hamiltonians use the
EM500 NN~\cite{EMPRC, EMrev1} and 3N (lnl)~\cite{Soma_LNL} interactions,
evolved by the similarity renormalization group to
$\lambda_{\rm SRG}=2.0\,\mathrm{fm}^{-1}$, with $\hbar\omega=20$~MeV.
The VQE calculations map time-reversed neutron pairs
to hard-core bosons~\cite{SYPRR2026}; all others use the Jordan--Wigner
encoding of Sec.~\ref{subsec:Encoding}.
\label{tab:instances}}
\begin{ruledtabular}
\begin{tabular}{>{\raggedright\arraybackslash}p{0.19\textwidth}>{\raggedright\arraybackslash}p{0.22\textwidth}>{\raggedright\arraybackslash}p{0.14\textwidth}>{\raggedright\arraybackslash}p{0.19\textwidth}>{\raggedright\arraybackslash}p{0.12\textwidth}}
\multicolumn{5}{l}{\textit{Resource estimates for encoded Hamiltonians (Secs.~\ref{sec:resource_estimation}, \ref{sec:Qresource_summary}, and \ref{sec:qpe_fixed_accuracy})}} \\[2pt]
Hamiltonian family & Interaction & Qubits $N_q$ & Truncation & Results \\
\hline
Valence: $p$, $sd$, $psd$, $pf$, $sd$--$pf$, $sd$--$pf$--$sdg$
 & CKpot, USDB, YSOX, GXPF1A, SDPF-MU, SDPFSDG-MU
 & 12--124
 & Inert core; one or more valence shells
 & Tables~\ref{tab:ModelSpace}, \ref{tab:Trotter_Error}, \ref{tab:measurement_grouping}; Figs.~\ref{fig:Trotter_error}--\ref{fig:QPE_fixed_accuracy} \\
No-core NN
 & EM500
 & 4--1,144
 & Structural counts, $e_{\max}\leq10$
 & Table~\ref{tab:ModelSpace}, Fig.~\ref{fig:Qresource_summary}(b) \\
 & & 4--880
 & Coefficient benchmarks, $e_{\max}\leq9$
 & Table~\ref{tab:Trotter_Error}, Figs.~\ref{fig:Trotter_error} and~\ref{fig:QPE_fixed_accuracy} \\
 & & 4--224
 & Greedy grouping, $e_{\max}\leq5$
 & Table~\ref{tab:measurement_grouping} \\
No-core NN+3N
 & EM500 + 3N (lnl)
 & 16--140
 & Structural counts and coefficient benchmarks, $e_{\max}\leq4$,
   $E_{3\max}=3e_{\max}$, no $N_{\max}$ truncation
 & Tables~\ref{tab:ModelSpace} and \ref{tab:Trotter_Error},
   Figs.~\ref{fig:Trotter_error}--\ref{fig:QPE_fixed_accuracy} \\
 & & 16--40
 & Greedy grouping, $e_{\max}=1,2$
 & Table~\ref{tab:measurement_grouping} \\
\hline\hline
\multicolumn{5}{l}{\textit{Statevector simulations of specific nuclei (Secs.~\ref{sec:finite_shot}, \ref{sec:State_prep}, and \ref{sec:demonstrations})}} \\[2pt]
Nucleus and space & Interaction & Algorithm & Compared with & Results \\
\hline
${}^{6}$He, ${}^{22}$O: neutron $p$ and $sd$ shells
 & CKpot, USDB
 & QPE, QKrylov, ODMD with finite shots
 & Exact ground-state energy in the $M_J=0$ sector (100~keV, 95\%)
 & Table~\ref{tab:r21_fixed_sampling} \\
${}^{3}$n: no-core, $e_{\max}=1$, $E_{3\max}=3$
 & EM500 + 3N (lnl)
 & QPE
 & Exact spectrum ($M=1/2$, $\pi=-$)
 & Fig.~\ref{fig:QPE_3n_e1} \\
${}^{6}$He: neutron $p$ shell
 & CKpot
 & QPE
 & Exact spectrum ($M=0$, $\pi=+$)
 & Fig.~\ref{fig:QPE_6He} \\
${}^{22}$O: neutron $sd$ shell
 & USDB
 & QKrylov, ODMD
 & Lowest five $J=0$ FCI energies
 & Fig.~\ref{fig:QKrylov_ODMD_22O} \\
${}^{18}$O: neutron $sd$ shell
 & USDB
 & Angular-momentum projection
 & $P(J=0)$ and ground-state overlap
 & Fig.~\ref{fig:state_prep_J0} \\
${}^{6}$He, ${}^{18}$O, ${}^{42}$Ca, ${}^{58}$Ni: two valence neutrons
 & CKpot, USDB, GXPF1A, JUN45
 & VQE (pUCCD)
 & DOCI and FCI energies
 & Appendix~\ref{app:vqe_demo} \\
\end{tabular}
\end{ruledtabular}
\end{table*}


\section{NuQuLib framework}
\label{sec:workflow}

Figure~\ref{fig:Sketch_Workflow} shows how
NuQuLib~\cite{Repo_NuQuLib,*Zenodo_NuQuLib} connects nuclear-structure
calculations to quantum circuits. The input consists of Hamiltonian
matrix elements in a specified model space. For example,
\texttt{NuHamil}~\cite{NuHamil_paper,*NuHamil_code} provides matrix
elements of two- and three-nucleon interactions derived from
ChEFT~\cite{Epelbaum2009,Machleidt2011,Epelbaum2020,EMrev1,EMrev2,LENPICrev1,LENPICrev2}.
These matrix elements also serve as inputs to nuclear many-body codes
such as \texttt{imsrg++}~\cite{imsrgcode}. For valence-space calculations,
NuQuLib can read phenomenological interactions in the \texttt{.snt}
format used by \texttt{KSHELL}~\cite{KSHELL1,*KSHELL2} and
\texttt{NuclearToolkit.jl}~\cite{NuclearToolkit.jl,*Repo_NuclearToolkit.jl}.

For the resource estimates in Sec.~\ref{sec:results}, we
use the standard Jordan--Wigner transformation to express the
second-quantized Hamiltonian as a weighted sum of Pauli strings.
This representation provides the operator terms and coefficients
that enter the circuit cost estimates.

For a specified nucleus and initial state, NuQuLib also provides
circuits for eigenvalue estimation. The classical processing depends
on the algorithm. QPE converts measured phases into energies, QKrylov
uses Hamiltonian and overlap matrix elements to solve a generalized
eigenvalue problem, and ODMD extracts energies from a time series of
overlaps. Section~\ref{sec:demonstrations} illustrates these calculations
for selected nuclear systems.
NuQuLib also supports variational workflows, in which a classical
optimizer updates circuit parameters, as indicated on the right-hand
side of Fig.~\ref{fig:Sketch_Workflow}. Appendix~\ref{app:vqe_demo}
gives a small example.

\subsection{Benchmark tasks and instances}
\label{subsec:benchmark_scope}

We use NuQuLib for three kinds of benchmark tasks, summarized in
Table~\ref{tab:instances}.
(i) Operator-level tests characterize encoded Hamiltonians in the full
qubit Hilbert space without targeting a specific nuclear state. They
comprise structural Pauli-term counts (Table~\ref{tab:ModelSpace}),
measurement-setting counts (Table~\ref{tab:measurement_grouping}),
coefficient norms and pairwise commutator coefficients
(Table~\ref{tab:Trotter_Error} and Fig.~\ref{fig:Trotter_error}), and
the QPE cost comparison at fixed energy accuracy
(Fig.~\ref{fig:QPE_fixed_accuracy}).
(ii) Finite-shot tests estimate the ground-state energies of
${}^{6}$He and ${}^{22}$O from specified initial determinants at a
common accuracy and success probability (Sec.~\ref{sec:finite_shot}).
(iii) Algorithm demonstrations simulate QPE, QKrylov, ODMD, and
angular-momentum projection for specified small systems with a
noise-free statevector simulator (Secs.~\ref{sec:State_prep} and
\ref{sec:demonstrations}).

The operator-level tests use both valence-space and no-core
Hamiltonians, which provide complementary settings. The no-core
formulation reflects the full complexity of the underlying interaction,
including the scaling associated with three-nucleon forces, whereas
valence-space Hamiltonians reach larger effective systems with fewer
active orbitals. Similar active-space constructions are used in
quantum chemistry~\cite{Veryazov:2011}. Considering both shows how
quantum resource requirements depend on approximations commonly
employed in nuclear structure theory.

Coefficient-screening thresholds define the retained operator, while
algorithmic energy tolerances describe accuracy relative to that operator.
Neither supplies a criterion for convergence of a physical nuclear
observable with respect to the interaction or model-space truncation.
The interaction inputs and numerical screening conventions are specified
in Appendix~\ref{app:pair_bound} and with the accompanying
data~\cite{Repo_Data}.

\section{Nuclear Hamiltonians, model spaces, and encoding}
\label{sec:Nuc_Hamiltonian}

\subsection{Nuclear Hamiltonians}
\label{subsec:hamiltonians}

A nucleus is governed by the strong interaction, but at the low energies
relevant for nuclear structure one can construct a Hamiltonian in terms
of nucleons. Because nucleons are composite, many-nucleon interactions
arise, for example through virtual excitations of the nucleon.
Three-nucleon interactions have been discussed since the pioneering work
by Fujita and Miyazawa~\cite{Fujita1957}, and their role in nuclei is an
active topic~\cite{Hammer2013,Hebeler2021,Coraggio2024}.
Including them, a general second-quantized nuclear Hamiltonian reads
\begin{align}
  H  = & \sum_{pq} h_{pq} a^\dagger_p a_q
  + \frac{1}{4} \sum_{pqrs} V_{pqrs} a^\dagger_p a^\dagger_q a_s a_r \nonumber \\
  & + \frac{1}{36} \sum_{pqrstu} V_{pqrstu} a^\dagger_p a^\dagger_q a^\dagger_r a_u a_t a_s \, ,
  \label{eq:Hamiltonian}
\end{align}
where $p=\left\{ n, l, j, j_{z}, t_{z} \right\}$ labels a single-particle
state by its principal quantum number $n$, orbital angular momentum $l$,
total angular momentum $j$, projection $j_z$, and isospin projection
$t_z$. The operator $a^{\dag}_{p}$ ($a_{p}$) creates (annihilates) a
nucleon in state $p$, and $h_{pq}$, $V_{pqrs}$, and $V_{pqrstu}$ are the
one-body, NN, and 3N matrix elements. The task is to solve
\begin{equation}
H | \Psi_{i} \rangle = E_{i} |\Psi_{i} \rangle
\end{equation}
for the eigenenergies $E_i$ and eigenstates $|\Psi_i\rangle$.

In the no-core shell model (NCSM)~\cite{Navratil2009}, all nucleons are
active and the Hamiltonian is diagonalized by configuration interaction
(CI). Classical applications are feasible for particle numbers
$A\lesssim20$~\cite{Barrett2013, Hergert2020}. In free space, $h_{pq}$ is
the kinetic energy, and $V_{pqrs}$ and $V_{pqrstu}$ can be derived from
ChEFT~\cite{Epelbaum2009, Machleidt2011}. This theory is built on chiral
symmetry and organizes the interactions by power
counting~\cite{Weinberg1990}. Its low-energy constants are fixed with
experimental data, many-nucleon interactions follow systematically, and
truncation uncertainties can be quantified~\cite{Melendez2017,
Melendez2019}. The matrix elements can be computed with the open-source
\texttt{NuHamil} code~\cite{NuHamil_paper}.

Valence-space CI instead divides the single-particle space into an inert
core, typically the orbitals of a doubly magic nucleus such as
$^{16}$O or $^{40}$Ca, an active space that carries the low-lying
degrees of freedom, and excluded higher-energy orbitals. This makes
heavier systems with $A>20$ accessible. With respect to the core, the
Hamiltonian becomes
\begin{equation}
\label{eq:H_valence}
H_{v} = \sum_{pq} f_{pq} :a^{\dag}_{p}a_{q}: + \frac{1}{4} \sum_{pqrs} \Gamma_{pqrs} :a^{\dag}_{p}a^{\dag}_{q}a_{s}a_{r}:\,,
\end{equation}
where the colons denote normal ordering with respect to the core,
$f_{pq}$ and $\Gamma_{pqrs}$ are one- and two-body matrix elements, and
the core energy is omitted. Three-body terms are usually neglected, as
most of their effect can be absorbed into $f_{pq}$ and $\Gamma_{pqrs}$.
These matrix elements can be derived from $H$ by
perturbative~\cite{Hjorth-Jensen1995} or
nonperturbative~\cite{Dikmen2015, Sun2018, Stroberg2019} decoupling of
the active space, fitted to experimental data, or obtained by combining
both approaches~\cite{Magilligan2021, Purcell2024, Kumar2025}. Examples
of fitted interactions are USD~\cite{Wildenthal1984, Brown1988, USDB}
for the $sd$ shell and KB3G~\cite{Poves2001} and
GXPF1~\cite{Honma2002, Honma2005} for the $pf$ shell; their matrix
elements are available in \texttt{KSHELL}~\cite{KSHELL1,*KSHELL2} and
\texttt{NuclearToolkit.jl}~\cite{NuclearToolkit.jl,*Repo_NuclearToolkit.jl}.
The structural term counts in Sec.~\ref{subsec:Encoding} depend only on
the model space, whereas the coefficient norms and commutator
coefficients used in Sec.~\ref{sec:results} depend on the interaction,
which we specify in each case.

\subsection{Model spaces and truncations}
\label{subsec:scaling}

To make the discussion concrete and to visualize difficulties in nuclear CI calculations, let us introduce the basis states.
A widely used representation is the so-called $M$-scheme,
where the many-body states are constructed by filling single-particle states with quantum numbers with a conserved
projection of the total angular momentum on the $z$-axis, $M$.
Another popular choice in nuclear structure calculations is the $J$-scheme representation,
where the many-body states are constructed with conserved total angular momentum $J$.

\begin{figure*}
\centering{
\includegraphics[width=0.95\textwidth]{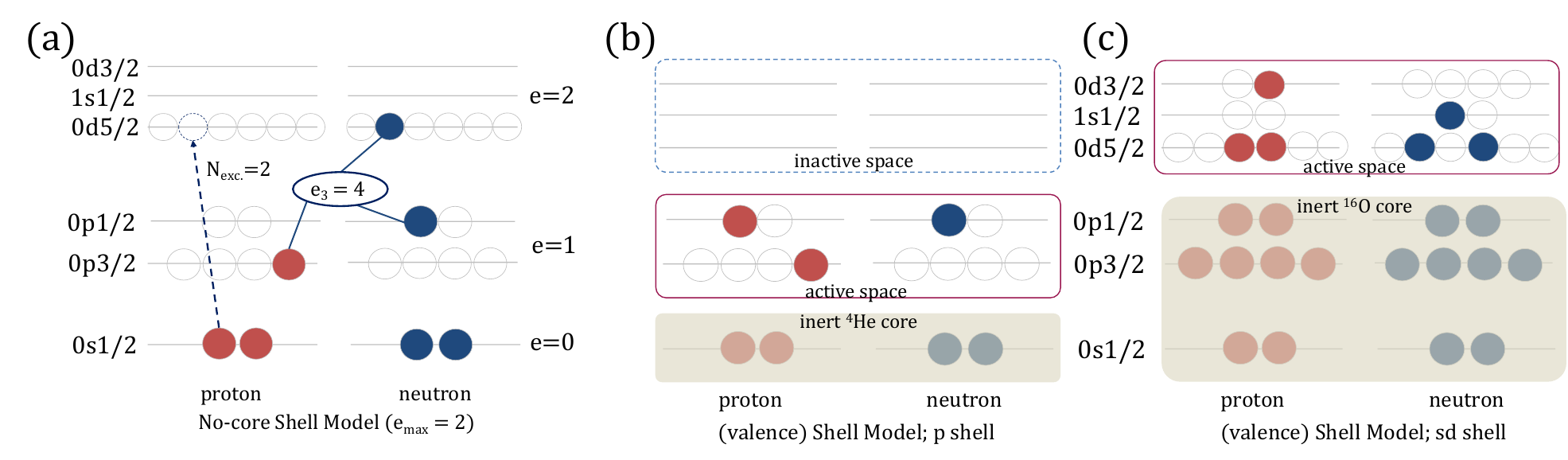}
\caption{Schematic illustration of nuclear configuration-interaction model spaces in the $M$-scheme representation.
The single-particle basis states are labeled by the quantum numbers $n$, $l$, $j$, $j_z$, and $t_z$.
Panel (a) shows a no-core shell-model space,
where all nucleons are active and the harmonic-oscillator single-particle basis
is truncated by ($e_{\max}$).
The figure also illustrates the commonly used three-body matrix-element cutoff $E_{3\max}$,
and a many-body excitation cutoff $N_\mathrm{max}$.
The label $e_3 =4$ denotes the sum of single-particle $e$ quanta associated with the three-body term shown schematically.
The arrow with the dashed line shows an example of excitations carrying $N_\mathrm{exc.}=2$.
Panels (b) and (c) show conventional valence-space shell-model calculations,
where an inert core is assumed and particle excitations are restricted to the chosen valence space.
\label{fig:Sketch_NuclCI}}
}
\end{figure*}

Figure~\ref{fig:Sketch_NuclCI} illustrates the model spaces in the
$M$-scheme. In the no-core example in panel~(a), all nucleons are
active~\cite{Navratil2009,Barrett2013}. The single-particle basis
contains harmonic-oscillator states with $e=2n+l\leq e_{\max}$.
For a three-body matrix element $V_{pqrstu}$ in
Eq.~\eqref{eq:Hamiltonian}, the cutoff $E_{3\max}$ requires both
$e_p+e_q+e_r\leq E_{3\max}$ and $e_s+e_t+e_u\leq E_{3\max}$.

A separate cutoff, $N_{\max}$, limits the total oscillator excitation
of a many-body configuration above the lowest allowed configuration
for the given nucleus. The dashed arrow in panel~(a) illustrates a
single-particle excitation of two oscillator quanta.
For the no-core NN+3N resource estimates, we set
$E_{3\max}=3e_{\max}$, so every triple of orbitals in the chosen
single-particle basis satisfies the three-body cutoff. We impose no
additional $N_{\max}$ restriction on the no-core resource estimates.

Panels~(b) and~(c) illustrate valence-space calculations with an
inert ${}^{4}$He core and an active $p$ shell, or an inert ${}^{16}$O
core and an active $sd$ shell, respectively~\cite{Brown2001,Caurier2005,Coraggio2009}.
Only the valence nucleons change configuration. As discussed above,
the interaction in the active space can be derived from an underlying
nuclear Hamiltonian or fitted to nuclear data.

\begin{figure}
  \centering
  \includegraphics[width=\linewidth]{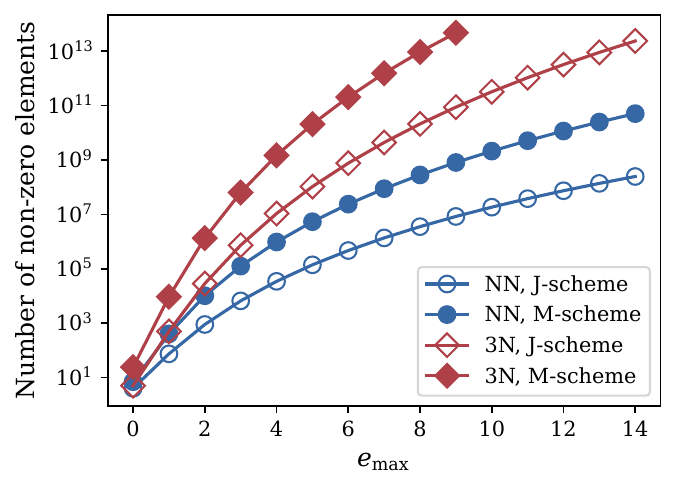}
  \caption{Numbers of two- and three-nucleon interaction matrix elements
  in the $J$- and $M$-scheme representations as a function of the
  single-particle cutoff $e_\mathrm{\max}$, where $2n+l\leq e_\mathrm{\max}$.
  The counts are obtained from the corresponding storage sizes using eight bytes per matrix element.
  }
  \label{fig:nonzero_ME_numbers}
\end{figure}

In Fig.~\ref{fig:nonzero_ME_numbers}, we show the number of non-zero matrix elements
in nuclear NN and 3N interactions within the $J$-scheme or $M$-scheme basis representations.
While the $J$-scheme representation is more compact due to the SU(2) symmetry reduction,
the $M$-scheme representation is more straightforward to implement and thus widely used in large-scale nuclear CI calculations.
For $N_q$ single-particle states, the numbers of two- and three-body
matrix elements are bounded by $O(N_q^4)$ and $O(N_q^6)$, respectively.
To quantify the growth in Fig.~\ref{fig:nonzero_ME_numbers}, we fit the
$M$-scheme counts to a power law in $N_q$, the number of proton and
neutron single-particle states at each $e_{\max}$.
The fitted exponents are 3.6 for NN over $0\leq e_{\max}\leq14$
and 5.4 for 3N over $0\leq e_{\max}\leq9$.
Already at $e_{\max}=6$, the $M$-scheme three-body matrix elements
occupy about 1.5~TiB at eight bytes per element.

\subsection{Jordan--Wigner encoding and Pauli-term counts}
\label{subsec:Encoding}

For the resource estimates, we use the Jordan--Wigner (JW)
transformation, with one qubit storing the occupation of each
single-particle state. We place proton states before neutron states
and order the magnetic substates of each orbital by increasing $j_z$.
For the structural counts in Table~\ref{tab:ModelSpace}, we enumerate
oscillator shells by increasing $e$ and orbitals within each shell by
increasing $j$. When reading a specific interaction, NuQuLib follows
the orbital order in the input file for each species.
For example, the oscillator ordering gives
\begin{align}
  \pi 0s_{1/2;j_z=-1/2}, & \pi 0s_{1/2; j_z=+1/2}, \pi 0p_{1/2; j_z=-1/2},
  \cdots, \nonumber \\
   \nu 0s_{1/2; j_z=-1/2},& \nu 0s_{1/2; j_z=+1/2}, \cdots,
\end{align}
where $\pi$ and $\nu$ denote proton and neutron states, respectively.
The ordering of single-particle states can be an important factor affecting
the efficiency of the mapping and the locality of the resulting qubit Hamiltonian.
Orderings chosen, for example, from the mutual information between qubits
have been proposed~\cite{Chiew2025OptimalFM,PhysRevB.109.115149,10946808,PRXQuantum.4.010326};
we do not optimize the ordering in this work.

As an example of JW-mapping, let us take the proton-neutron two-body interaction term $V_{pqrs} \pi^\dagger_p \nu^\dagger_q \pi_r \nu_s$.
We represent protons and neutrons in separate Fock spaces, with fermionic anticommutation relations
within each species and commuting operators between species. For $p<r$ and $q<s$, the JW mapping gives
\begin{align}
\pi^{\dagger}_p \pi_r & \otimes  \nu^{\dagger}_q  \nu_s \nonumber \\
  \mapsto &
\left \{ \frac{1}{4} 
(X^{\pi}_p - iY^{\pi}_p)\otimes Z^{\pi}_{p+1} \otimes \cdots \otimes (X^{\pi}_r + iY^{\pi}_r) 
\right\}
 \nonumber \\
& \otimes \left \{ \frac{1}{4}
\left( X^{\nu}_q - iY^{\nu}_q \right) \otimes Z^{\nu}_{q+1} \otimes \cdots \otimes (X^{\nu}_s + iY^{\nu}_s)
\right\},
\end{align}
where $\pi$ and $\nu$ denote proton and neutron operators, respectively.

Alternative encodings and orbital orderings can improve Pauli locality:
JW strings can have weight $O(N_q)$, whereas Bravyi--Kitaev (BK)
strings have weight $O(\log N_q)$ for fixed-body-rank operators~\cite{Steudtner2018Mappings}.
Their impact depends on the resource metric. Without symmetry reduction,
standard JW, parity, and BK encodings are related by Clifford conjugation,
which preserves Pauli-term counts, coefficient magnitudes, and commutation relations.
Thus, these encodings give the same number $M$ of retained
nonidentity Pauli terms and the same sum $\lambda'$ of their
coefficient magnitudes. They also give the same pairwise commutator
coefficient $C_{\rm pair}$ in the first-order Trotter error bound
of Sec.~\ref{sec:pairwise_bounds}.
With corresponding product-formula term orderings,
the operator-norm Trotter error is also unchanged; independently optimized
orderings need not have the same error. Shorter strings can reduce CNOT
costs and circuit depth, but do not by themselves reduce the termwise
rotation-based T-gate count used here. Fully commuting grouping has the same
commutation graph, whereas qubit-wise commuting grouping can change.
Symmetry tapering and compact encodings of restricted configuration
spaces, which can reduce qubit counts, are discussed in
Sec.~\ref{subsec:extensions}.

For the charge-conserving Hamiltonians considered here, number conservation
requires an even number of Pauli-$X/Y$ factors in each nucleon species.
Hermiticity requires real Pauli coefficients, while the reality of the
Hamiltonian matrix in the adopted occupation basis excludes strings with
an odd total number of Pauli-$Y$ factors.
The number of $Y$ factors need not be even in each species separately:
terms containing $X_iY_j$ in both proton and neutron sectors are allowed.
These restrictions also apply to 3N terms. In addition, we retain only
combinations of creation and annihilation operators that conserve the
total angular-momentum projection and parity.

Table~\ref{tab:ModelSpace} lists the number of system qubits $N_q$ and
the number of distinct Jordan--Wigner Pauli strings $N_{\hat H}$,
including the identity, for each model space. We count all strings
allowed by the operator structure and symmetry selection rules,
without using the coefficients of a particular interaction. This gives
an upper bound on the term count for a specific Hamiltonian, since
some coefficients can vanish or fall below the numerical cutoff.
For the 20 benchmark Hamiltonians we checked, the tabulated counts
were within $0.3\%$ of the counts retained after coefficient screening.
We therefore use the tabulated counts to estimate how circuit costs
grow with model-space size at fixed algorithm parameters.
For the no-core spaces, fits of $N_{\hat H}$ in
Table~\ref{tab:ModelSpace} to a power law in $N_q$ give exponents of
about 3.7 for NN and 6.1 for NN+3N, comparable to those of the
matrix-element counts in Fig.~\ref{fig:nonzero_ME_numbers}. The growth of
the encoded Hamiltonian, and thus of the cost of one product-formula
step, follows directly from the growth of the interaction.

\begin{table}[t]
\centering
\caption{Model spaces, system-qubit counts $N_q$, and counts $N_{\hat{H}}$ of distinct Jordan--Wigner Pauli strings used in the resource estimates.
Counts include one-body contributions and the identity, with all proton and neutron interaction channels included.
They follow from the operator structure and symmetry selection rules; interaction-specific coefficient zeros and cancellations are not evaluated.
The NN+3N entries use $E_\mathrm{3max}=3e_\mathrm{max}$.
\label{tab:ModelSpace}}
\begin{ruledtabular}
\begin{tabular}{lrr}
Model space & $N_q$ & $N_{\hat{H}}$ \\
\hline
\multicolumn{3}{c}{\textit{Valence, NN}} \\
$p$ shell & $12$ & $975$ \\
$sd$ shell & $24$ & $12\,869$ \\
$pf$ shell & $40$ & $82\,861$ \\
$psd$ shell & $36$ & $36\,563$ \\
$sd$--$pf$ shell & $64$ & $299\,025$ \\
$sd$--$pf$--$sdg$ & $124$ & $3\,764\,359$ \\
\hline
\multicolumn{3}{c}{\textit{No-core, NN}} \\
$e_\mathrm{max}=0$ & $4$ & $19$ \\
$e_\mathrm{max}=1$ & $16$ & $1\,793$ \\
$e_\mathrm{max}=2$ & $40$ & $57\,445$ \\
$e_\mathrm{max}=3$ & $80$ & $767\,065$ \\
$e_\mathrm{max}=4$ & $140$ & $6\,182\,807$ \\
$e_\mathrm{max}=5$ & $224$ & $35\,474\,065$ \\
$e_\mathrm{max}=6$ & $336$ & $159\,654\,305$ \\
$e_\mathrm{max}=7$ & $480$ & $598\,371\,049$ \\
$e_\mathrm{max}=8$ & $660$ & $1\,944\,113\,939$ \\
$e_\mathrm{max}=9$ & $880$ & $5\,631\,098\,241$ \\
$e_\mathrm{max}=10$ & $1\,144$ & $14\,842\,952\,717$ \\
\hline
\multicolumn{3}{c}{\textit{No-core, NN+3N}} \\
$e_\mathrm{max}=1$ & $16$ & $51\,377$ \\
$e_\mathrm{max}=2$ & $40$ & $15\,346\,373$ \\
$e_\mathrm{max}=3$ & $80$ & $980\,779\,057$ \\
$e_\mathrm{max}=4$ & $140$ & $26\,654\,802\,915$ \\
\end{tabular}
\end{ruledtabular}
\end{table}


\section{Quantum Algorithms for Eigenvalue Problems}
\label{sec:quantum_algorithms}

\begin{figure}[htbp]
  \centering
  \includegraphics[width=\linewidth]{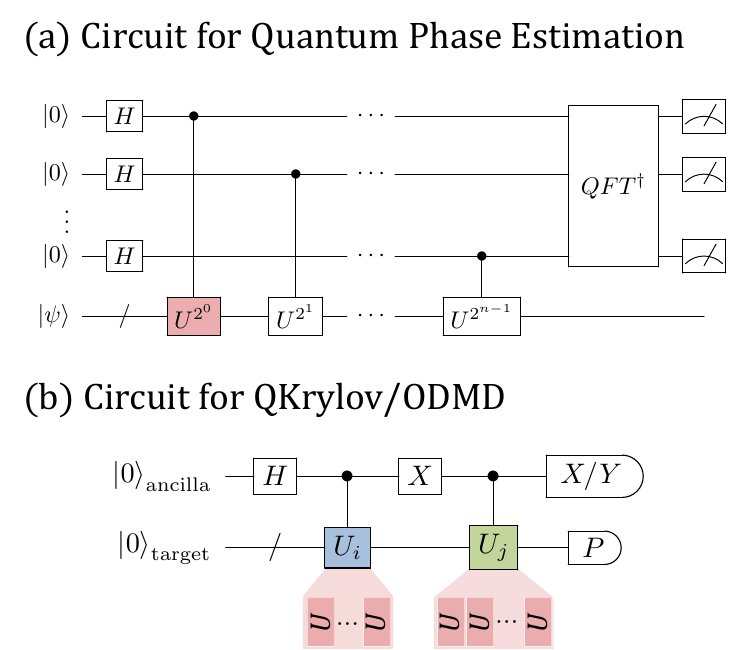}
  \caption{Quantum circuits for the algorithms considered in this work.
(a) Quantum Phase Estimation (QPE): controlled time evolution is followed by the inverse quantum Fourier transform.
(b) Quantum Krylov Subspace Method (QKrylov)/Observable Dynamic Mode Decomposition (ODMD): interference between time-evolved states gives overlaps or Pauli matrix elements.
Red blocks denote the basic evolution $U=\exp(-iH\delta t)$; the highlighted block in (a) is $U^{2^0}=U$.
Blue and green blocks in (b) prepare the two time-evolved states, and the lower red blocks show their repeated applications of $U$.
\label{fig:Circuits}
  }
\end{figure}

We consider three methods for estimating eigenenergies of a qubit
Hamiltonian $H$ from an initial state $\ket{\Phi_0}$ with nonzero
overlap with the eigenstate of interest.
Figure~\ref{fig:Circuits} shows how the methods use time evolution
under $H$ to obtain spectral information. We describe the measurements
and classical processing for each method below.
Section~\ref{sec:resource_estimation} defines the cost model, and
Sec.~\ref{sec:results} gives the resulting resource estimates.

QPE~\cite{Kitaev95} applies controlled powers of the evolution
operator $U=e^{-iH\delta t}$ and reads out an eigenphase from the
ancilla register after an inverse quantum Fourier transform.
QKrylov~\cite{Cortes_PRA22} uses time-evolved states as a basis.
It measures their overlaps and Hamiltonian matrix elements, then
solves the resulting generalized eigenvalue problem classically.
ODMD~\cite{ODMD_Shen} samples the overlap
$\langle\Phi_0|e^{-iHt}|\Phi_0\rangle$ at a sequence of times and
extracts energies from the frequencies present in this signal.

\subsection{Quantum Phase Estimation}

In Fig.~\ref{fig:Circuits}(a), QPE estimates an eigenphase of the
time-evolution operator $U=e^{-iH\delta t}$. For a Hamiltonian
eigenstate $\ket{\Psi_j}$ with energy $E_j$, we have
\begin{equation}
  U\ket{\Psi_j}=e^{2\pi i\phi_j}\ket{\Psi_j},
  \qquad \phi_j=-\frac{E_j\delta t}{2\pi}\pmod{1}.
\end{equation}
After the inverse quantum Fourier transform, measuring the $N_a$ phase
qubits returns an integer $m$ such that $m/2^{N_a}$ estimates $\phi_j$.
Energies separated by $2\pi/|\delta t|$ give the same phase, so
converting the readout to an energy requires specifying the energy
interval of interest. Within that interval, adjacent readout bins are
separated by $2\pi/(2^{N_a}|\delta t|)$. At fixed $\delta t$, each
additional phase qubit halves this spacing and doubles the largest
controlled power of $U$ in the circuit.

For an initial state $\ket{\Phi_0}=\sum_j c_j\ket{\Psi_j}$, each
eigenstate contributes to the readout distribution with weight
$|c_j|^2$. A small ground-state weight can therefore require many
shots to resolve the ground-state peak even when the energy bins are
sufficiently narrow.

The resource estimates use this phase-register readout for both
Trotterized and qubitized QPE. In qubitized QPE, the controlled
unitary is a walk operator, and the energy is recovered from the
cosine of its eigenphase, as described in
Appendix~\ref{app:qubitized_QPE}.
Other phase-estimation protocols, such as Bayesian approaches~\cite{Wiebe_PRL16,Yamamoto_PRR24}
and robust phase estimation~\cite{Kimmel_PRA15,Rudinger_PRL17}, are not considered here.

\subsection{Quantum Krylov Subspace Method}

QKrylov estimates low-lying energies by diagonalizing the Hamiltonian
in a subspace spanned by time-evolved states~\cite{Cortes_PRA22}.
Starting from an initial state $\ket{\Phi_0}$, we construct basis states
$\ket{\Phi_k}=e^{-iHt_k}\ket{\Phi_0}$ at a set of equally spaced
times $t_k$. We approximate an eigenstate in this subspace by
$\ket{\Psi}=\sum_k c_k\ket{\Phi_k}$.
Since the basis states are generally nonorthogonal, their expansion
coefficients $\mathbf{c}$ satisfy the generalized eigenvalue problem
\begin{align}
  \tilde{H}\mathbf{c} &= E\tilde{N}\mathbf{c}, \label{eq:QKrylov}\\
  \tilde{H}_{kl} &= \bra{\Phi_k}H\ket{\Phi_l}, \label{eq:Hmatrix}\\
  \tilde{N}_{kl} &= \langle\Phi_k|\Phi_l\rangle. \label{eq:Nmatrix}
\end{align}
The quantum circuits estimate the entries of $\tilde{H}$ and
$\tilde{N}$. Classical diagonalization then gives the approximate
energies and the corresponding coefficient vectors.

To estimate $\tilde{H}_{kl}$, we group Pauli terms that can be
measured in the same setting. For an off-diagonal element, ancilla
measurements in the $X$ and $Y$ bases give its real and imaginary
parts. The same measurement outcomes also give the overlap
$\tilde{N}_{kl}$ through the ancilla marginal.

Grouping determines how many distinct measurement settings are
needed. The number of shots per setting also depends on the target
energy accuracy. In particular, when the Krylov states are nearly
linearly dependent, small statistical errors in the measured matrices
can produce appreciable errors in the estimated energies.

With exact time evolution on a uniform time grid, both
$\tilde{N}_{kl}$ and $\tilde{H}_{kl}$ depend only on $l-k$ and
therefore have Toeplitz structure. Yoshioka et al.~\cite{Yoshioka_npjQ25}
used this structure to simplify the measurement circuits in a hardware
demonstration. Repeating a fixed Trotter step preserves the Toeplitz
structure of $\tilde{N}$, but the step need not commute with $H$,
so $\tilde{H}$ need not be Toeplitz. Our resource estimates count
the measurements for each distinct pair of Krylov states separately.

\subsection{Observable Dynamic Mode Decomposition}

ODMD~\cite{ODMD_Shen} applies dynamic mode decomposition
(DMD)~\cite{DMD_Schmid,Rowley2009} to a time series of quantum
measurements. Here the signal is the autocorrelation of the initial
state. For $\ket{\Phi_0}=\sum_j c_j\ket{\Psi_j}$ and
$t_k=k\delta t$, it is
\begin{equation}
  o_k \equiv \langle\Phi_0|e^{-iHt_k}|\Phi_0\rangle
      = \sum_j |c_j|^2 e^{-iE_jt_k}.
\end{equation}
Thus an eigenstate contributes an oscillation whose frequency encodes
its energy and whose weight is its squared overlap with the initial
state.

For a normalized initial state, $o_0=1$ is known. We sample the real
and imaginary parts of $o_k$ at $K=N_{\mathrm{snap}}$ nonzero times
using two ancilla measurement settings per time. The number of shots
per setting determines the statistical precision of each component.

From the sequence $o_0,\ldots,o_K$, we form two Hankel matrices,
each with $d$ rows and $K-d+1$ columns,
\begin{equation}
\begin{aligned}
X &= \left[
\begin{array}{cccc}
  o_0     & o_1    & \cdots & o_{K-d}   \\
  o_1     & o_2    & \cdots & o_{K-d+1} \\
  \vdots  & \vdots &        & \vdots    \\
  o_{d-1} & o_d    & \cdots & o_{K-1}
\end{array}
\right], \\[4pt]
Y &= \left[
\begin{array}{cccc}
  o_1    & o_2     & \cdots & o_{K-d+1} \\
  o_2    & o_3     & \cdots & o_{K-d+2} \\
  \vdots & \vdots  &        & \vdots    \\
  o_d    & o_{d+1} & \cdots & o_K
\end{array}
\right].
\end{aligned}
\label{eq:Hankel}
\end{equation}
Each column contains $d$ consecutive samples, and each anti-diagonal
reuses a single value of $o_k$. The columns of $Y$ are shifted forward
by one time step relative to the corresponding columns of $X$.

We fit a linear map $A$ such that $Y\simeq AX$, using a truncated
singular value decomposition (SVD) of $X$. Discarding small singular
values limits the amplification of sampling noise in the fit.
For a resolved mode, an eigenvalue $z_j$ of $A$ approximates
$e^{-iE_j\delta t}$. We therefore estimate its energy from
$-\arg(z_j)/\delta t$ on the chosen energy branch, with the same
period $2\pi/|\delta t|$ as in time-evolution QPE.

Extensions include multi-observable ODMD~\cite{Shen:2026} and Fourier-denoised ODMD
~\cite{Bassi:2025}.

\subsection{Remarks on State Preparation and Algorithmic Trade-offs}

The ground-state weight $|\langle\Psi_0|\Phi_0\rangle|^2$ enters the
three algorithms in different ways. In QPE it gives the weight of the
ground-state contribution to the phase readout. In ODMD it is the
coefficient of the ground-state oscillation in $o_k$, so a weak
component can be difficult to distinguish from sampling noise. In
QKrylov, the initial state determines which eigenstates are represented
in the time-evolved basis. If the ground-state overlap vanishes, exact
time evolution cannot introduce a ground-state component into that basis.

Each shot requires a fresh preparation of the nuclear input state.
For a given preparation circuit, its total cost therefore scales with
the total number of shots across all measurement settings.
The reported T-gate counts exclude nuclear state preparation
(Sec.~\ref{sec:assumptions}); Sec.~\ref{sec:State_prep} discusses possible
preparation methods and their overheads, including a literature example
of the cost and ground-state overlap.

\section{Cost model}
\label{sec:resource_estimation}

This section specifies the cost model used for the resource estimates
in Sec.~\ref{sec:results}. We first state the assumptions and summarize
all parameters in Table~\ref{tab:cost_model}. We then define the
pairwise commutator coefficient that controls the first-order Trotter
error, describe the grouping of Pauli terms into measurement settings,
and give the T-gate count formulas.

\subsection{Scope and assumptions}
\label{sec:assumptions}

We use the Jordan--Wigner encoding with one system qubit per
single-particle state. The valence-space Hamiltonians contain up to
two-body terms. The no-core estimates include both NN and NN+3N cases.
For NN+3N we set $E_{3\max}=3e_{\max}$, retaining all three-body
matrix elements within the chosen single-particle space. The no-core
estimates impose no additional $N_{\max}$ truncation.

For the scaling estimates at fixed algorithm parameters, we use the
structural counts $N_{\hat H}$ in Table~\ref{tab:ModelSpace}. These
include the identity and count the Pauli strings allowed by the
operator structure and symmetry selection rules described in
Sec.~\ref{subsec:Encoding}. For the measurement-grouping benchmarks
and the QPE comparison at fixed energy accuracy, we use the nonidentity
terms retained after coefficient screening for each Hamiltonian.

We assume that the nuclear input state has already been prepared and count the cost of the evolution and measurement circuits. Nuclear state preparation is excluded, and its
efficiency is not established for the full set of model spaces.
A complete ground-state energy estimate must also account for preparation
on each circuit execution and the effect of target-state overlap on
algorithm-specific sampling requirements. Section~\ref{sec:State_prep}
gives a literature-based example of the preparation cost and overlap.

In the fixed-register resource estimates in Fig.~\ref{fig:Qresource_summary}(a,b),
we represent one application of $U=\exp(-iH\delta t)$ by one complete
first-order product-formula step.
The powers $U^{2^k}$ are implemented by repeating this basic circuit $2^k$ times.
Appendix~\ref{app:qubitized_QPE} instead determines the number of steps per
basic evolution from a specified energy-error budget and compares the
resulting cost with qubitized QPE.

\begin{figure}
  \centering
  \includegraphics[width=\linewidth]{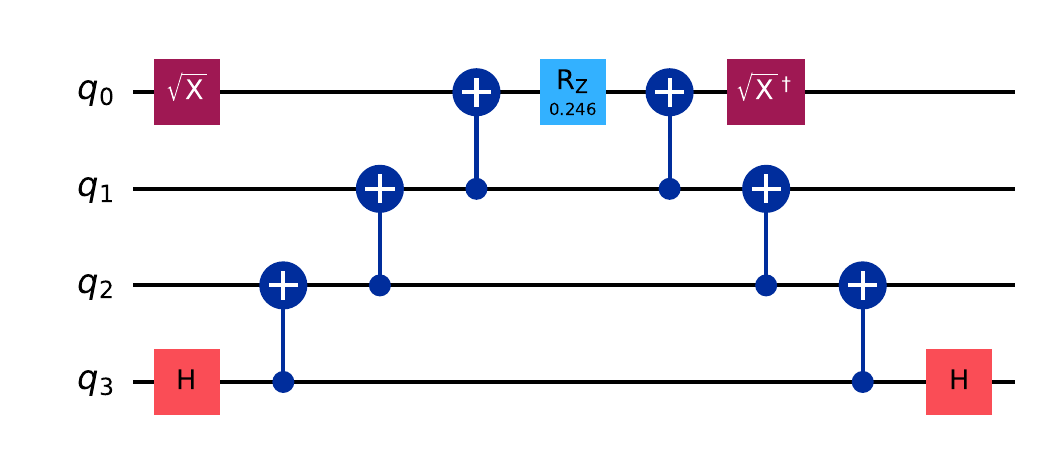}
  \caption{
Circuit for the Pauli-string exponential $\exp[-i\theta Y_0 Z_1 Z_2 X_3]$ with $\theta=0.123$.
The central gate is $R_z(2\theta)$ on qubit $q_0$.
}
\label{fig:time_evolution_circuit}
\end{figure}

In the product formula, we implement each Pauli-string exponential
separately. Figure~\ref{fig:time_evolution_circuit} illustrates the
uncontrolled circuit. Single-qubit Clifford gates change the local
Pauli bases, and a CNOT ladder accumulates the parity on one qubit.
After a single $R_z$ rotation, the circuit reverses the ladder and
the basis changes. We count each term separately, without cancelling
gates between successive terms.

Our primary resource metric is the logical T-gate count
~\cite{Bravyi:2005,Hao:2026}. We estimate the cost of synthesizing an
$R_z$ rotation to precision $\epsilon$ using the leading expression
$T_\epsilon \simeq 3\log_2(1/\epsilon)$~\cite{Ross_Selinger_2016}.
For the fixed-parameter estimates in Fig.~\ref{fig:Qresource_summary}(a,b),
we take $T_\epsilon=100$, corresponding to $\epsilon\sim10^{-10}$
in this model~\cite{NERSC25}.

An uncontrolled Pauli-string exponential uses one synthesized rotation.
Our decomposition of its controlled version uses two. Thus, for one
first-order Trotter step in panels~(a,b), we count
\[
  T_{\hat U}=(N_{\hat H}-1)T_\epsilon, \qquad
  T_{c\hat U}=2T_{\hat U}.
\]
We omit the identity term from the circuit by evolving with
$H'=H-h_I I$ and adding the known offset $h_I$ to the estimated energy.

The QPE comparison at fixed energy accuracy (Fig.~\ref{fig:QPE_fixed_accuracy}) uses the same identity shift and the coefficients of each
retained Hamiltonian. For each QPE circuit with $n_R$ synthesized
rotations, we assign an operator-norm error of at most $10^{-3}/n_R$
to each rotation, so their combined contribution to the circuit error
is at most $10^{-3}$. Section~\ref{sec:cost_formulas} gives the
corresponding T-gate counts at the common energy accuracy.

QKrylov solves the generalized eigenvalue problem in
Eq.~\eqref{eq:QKrylov}. ODMD uses a truncated SVD of $X$ to fit
the map from $X$ to $Y$ in Eq.~\eqref{eq:Hankel} and extracts
energies from the eigenphases of the fitted map. The logical T-gate counts
cover the quantum circuits only.
Table~\ref{tab:cost_model} collects the settings of the three sets of
results in Sec.~\ref{sec:results}. Section~\ref{subsec:limitations}
summarizes the limitations of these cost models.

\begin{table*}
\caption{Summary of the cost models. Figures~\ref{fig:Qresource_summary} and
\ref{fig:QPE_fixed_accuracy} report logical T-gate counts excluding nuclear
state preparation, Clifford gates, and error correction. Table~\ref{tab:r21_fixed_sampling}
reports shot counts and compares the weighted number $W$ of time-evolution
Pauli rotations over all shots, before gate synthesis.
\label{tab:cost_model}}
\begin{ruledtabular}
\begin{tabular}{>{\raggedright\arraybackslash}p{0.13\textwidth}>{\raggedright\arraybackslash}p{0.25\textwidth}>{\raggedright\arraybackslash}p{0.25\textwidth}>{\raggedright\arraybackslash}p{0.25\textwidth}}
 & Fixed algorithm parameters [Fig.~\ref{fig:Qresource_summary}(a,b)]
 & QPE at fixed energy accuracy (Fig.~\ref{fig:QPE_fixed_accuracy})
 & Finite-shot estimates (Table~\ref{tab:r21_fixed_sampling}) \\
\hline
Algorithms & Trotterized QPE, QKrylov, ODMD & Trotterized and qubitized QPE & QPE, QKrylov, ODMD \\
Hamiltonian terms & Structural count $N_{\hat H}-1$ (Table~\ref{tab:ModelSpace}) & Retained count $M$ (Table~\ref{tab:Trotter_Error}) & Retained neutron-space terms (105 for ${}^{6}$He, 1,690 for ${}^{22}$O) \\
Time evolution & First order, $r=1$ per interval & First order with $r$ from $C_{\rm pair}$ [Eq.~\eqref{eq:qpe_common_steps}]; qubitized walk & First order, $r=4$ per interval \\
Basic interval $\delta t$ & $0.01\,\mathrm{MeV}^{-1}$ & $\pi/(2\lambda')$ & Chosen per method (Table~\ref{tab:r21_fixed_sampling}) \\
Algorithm sizes & $N_a=10$, 20; $N_{\rm iter}=50$; $N_{\rm snap}=50$ & $N_a=m_T$ or $m_W$ [Eq.~\eqref{eq:qpe_common_bits}] & Table~\ref{tab:r21_fixed_sampling} \\
Rotation synthesis & $T_\epsilon=100$ per rotation & Circuit budget $10^{-3}$ in operator norm & Not synthesized; rotation work $W$ \\
QKrylov settings & Majorana cover bound, Eq.~\eqref{eq:grouping_cost_prescription} & --- & Greedy Majorana groups ($G=16$ and 114) \\
Repetitions & $N_{\rm shot}=1$ per circuit & One spectral sample & Calibrated (Appendix~\ref{app:r21_fixed_sampling}) \\
Accuracy target & None common to the methods & $\Delta E=100$~keV (also 1~MeV and 1~keV). Phase readout and Hamiltonian approximation each receive $\Delta E/2$ (50~keV at the 100~keV target) & $100$~keV with success probability $\geq95\%$ \\
\end{tabular}
\end{ruledtabular}
\end{table*}

\begin{table*}
\centering
\caption{Retained nonidentity term counts $M$, coefficient norms
$\lambda'$, and pairwise commutator coefficients $C_{\rm pair}$ of the
benchmark Hamiltonians, with $\lambda'$ in MeV and $C_{\rm pair}$ in
$\mathrm{MeV}^2$. The identity does not contribute. No evolution time is
required to define $C_{\rm pair}$. The NN+3N rows use
$E_{3\max}=3e_{\max}$. Input conventions are specified in
Appendix~\ref{app:pair_bound}.
\label{tab:Trotter_Error}}
\begin{ruledtabular}
\begin{tabular}{lrrrr}
Hamiltonian & $N_q$ & $M$ & $\lambda'$ & $\log_{10}C_{\rm pair}$\\
\hline
\multicolumn{5}{c}{\textit{Valence}} \\
CKpot ($p$) & 12 & 974 & 238.6 & 3.98 \\
USDB ($sd$) & 24 & 12,868 & 962.7 & 5.02 \\
YSOX ($psd$) & 36 & 36,562 & 2,631 & 5.79 \\
GXPF1A ($pf$) & 40 & 82,860 & 2,123 & 5.54 \\
SDPF-MU ($sd$--$pf$) & 64 & 299,024 & 8,705 & 6.62 \\
SDPFSDG-MU ($sd$--$pf$--$sdg$) & 124 & 3,763,622 & $6.32\times10^{4}$ & 8.10 \\
\hline
\multicolumn{5}{c}{\textit{No-core, NN}} \\
$e_{\max}=0$ & 4 & 18 & 28.9 & 1.29 \\
$e_{\max}=1$ & 16 & 1,792 & 631.6 & 4.81 \\
$e_{\max}=2$ & 40 & 57,284 & 6,507 & 6.57 \\
$e_{\max}=3$ & 80 & 766,680 & $3.92\times10^{4}$ & 7.87 \\
$e_{\max}=4$ & 140 & 6,180,598 & $1.69\times10^{5}$ & 8.92 \\
$e_{\max}=5$ & 224 & 35,470,272 & $5.77\times10^{5}$ & 9.79 \\
$e_{\max}=6$ & 336 & 159,642,144 & $1.66\times10^{6}$ & 10.54 \\
$e_{\max}=7$ & 480 & 598,352,536 & $4.22\times10^{6}$ & 11.19 \\
$e_{\max}=8$ & 660 & 1,944,065,842 & $9.70\times10^{6}$ & 11.78 \\
$e_{\max}=9$ & 880 & 5,631,018,552 & $2.06\times10^{7}$ & 12.31 \\
\hline
\multicolumn{5}{c}{\textit{No-core, NN+3N}} \\
$e_{\max}=1$ & 16 & 51,376 & 738.6 & 5.00 \\
$e_{\max}=2$ & 40 & 15,343,844 & $1.54\times10^{4}$ & 7.51 \\
$e_{\max}=3$ & 80 & 980,709,976 & $2.31\times10^{5}$ & 9.69 \\
$e_{\max}=4$ & 140 & 26,651,010,918 & $2.23\times10^{6}$ & 11.47 \\
\end{tabular}
\end{ruledtabular}
\end{table*}

\subsection{Trotter error and the pairwise commutator coefficient}
\label{sec:pairwise_bounds}

With $\hbar=1$, we use the identity-shifted Hamiltonian
$H'=H-h_I I=\sum_{j=1}^{M}H_j$, where $H_j=h_jP_j$ and the $P_j$
are distinct nonidentity Pauli strings. The retained coefficients
$h_j$ are in MeV. The known shift $h_I$ is restored when reporting
energies. For a first-order product formula, define
\begin{align}
S_1(\tau)&=\prod_{j=1}^{M}e^{-i\tau H_j},\\
\epsilon_{\rm PF}(t,r)&=\left\|S_1(t/r)^r-e^{-it\sum_jH_j}\right\|,\\
C_{\rm pair}&=\frac12\sum_{j<k}\|[H_j,H_k]\|\nonumber\\
&=\sum_{\substack{j<k\\\{P_j,P_k\}=0}}|h_jh_k|,\\
\epsilon_{\rm PF}(t,r)&\le B_{\rm pair}(t,r)=\frac{t^2}{r}C_{\rm pair}.\label{eq:Bound_Pair}
\end{align}
The norm is the operator norm on the full qubit Hilbert space. The bound follows
from the first-order commutator bound, the triangle inequality, and telescoping over
$r$ unitary steps~\cite{Childs:2021}.
For the same decomposition into individual Pauli terms, the
order-independent commutator sum $\alpha$ in Eq.~(6) of
Ref.~\cite{MartinezMartinez2023} equals $4C_{\rm pair}$.
$C_{\rm pair}$ has units of $\mathrm{MeV}^2$;
$t$ has units of $\mathrm{MeV}^{-1}$; both $\epsilon_{\rm PF}$ and $B_{\rm pair}$ are dimensionless.

We evaluate $C_{\rm pair}$ from the coefficients of all Hamiltonian terms. In the Majorana representation, two terms
anticommute when their index sets overlap in an odd number of
indices. Appendix~\ref{app:pair_bound} uses this property to sum
the weighted contributions without enumerating every pair.
The resulting coefficient is independent of $t$ and $r$. The bound
$B_{\rm pair}(t,r)$ can overestimate the actual evolution error
because it sums commutator norms and discards cancellations.
For a target simulation error $0<\epsilon_{\rm sim}<2$, a sufficient condition is
\begin{equation}
r\ge\max\!\left(1,\left\lceil
\frac{C_{\rm pair}t^2}{\epsilon_{\rm sim}}\right\rceil\right).
\end{equation}
This is a conservative guarantee, not the minimum number of steps required in practice.
For choices of $t,r$ with $B_{\rm pair}(t,r)\ge2$, the estimate does not improve on
the trivial unitary-distance bound of 2 and gives no useful error guarantee for that choice.
For Fig.~\ref{fig:QPE_fixed_accuracy}, we apply the bound over
one basic interval $\delta t$. Equation~\eqref{eq:qpe_common_steps}
converts the $50$~keV allowance for the Trotter spectral error into
the dimensionless tolerance used to choose $r$.

Table~\ref{tab:Trotter_Error} lists, for each retained Hamiltonian, the
nonidentity term count $M$, the coefficient norm
$\lambda'=\sum_{j=1}^{M}|h_j|$, and $C_{\rm pair}$. These are the inputs
to the QPE comparison at fixed energy accuracy in
Sec.~\ref{sec:qpe_fixed_accuracy}.

\begin{figure}[htbp]
\centering
\includegraphics[width=\linewidth]{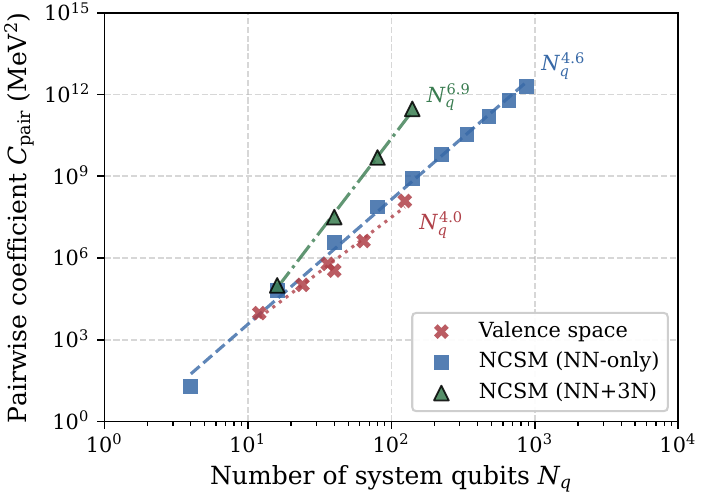}
\caption{Pairwise commutator coefficients $C_{\rm pair}$ for the 20 benchmark Hamiltonians in Table~\ref{tab:Trotter_Error}. Lines show separate unweighted least-squares fits in log--log space
for each family, restricted to its evaluated range. The displayed exponents describe
finite-range trends.
\label{fig:Trotter_error}}
\end{figure}

Table~\ref{tab:Trotter_Error} and Fig.~\ref{fig:Trotter_error} show $C_{\rm pair}$ for six valence-space
Hamiltonians, ten no-core NN Hamiltonians, and four no-core NN+3N Hamiltonians.
At $e_{\max}=2$, adding the 3N interaction increases the structural
Pauli-term count by a factor of about 267 (Table~\ref{tab:ModelSpace}). For these two Hamiltonians in the same model space,
the coefficient $C_{\rm pair}$ increases by a factor of about 8.6,
since each anticommuting pair is weighted by $|h_jh_k|$.

Fits of $C_{\rm pair}$ versus $N_q$ give exponents of 4.0, 4.6,
and 6.9 for the valence, no-core NN, and no-core NN+3N series,
respectively. These exponents describe the sampled ranges. The
NN+3N fit uses only four points, and restricting the no-core NN
fit to $N_q\geq80$ changes its exponent from 4.6 to 4.3.

\begin{table*}[!t]
\caption{Achieved measurement-setting reductions $R_{\mathrm{set}}=M/G$
for the same retained Hamiltonian terms, with nonidentity count $M$.
A dash marks an unfinished greedy calculation. Input conditions and
limitations are described in Sec.~\ref{sec:measurement_grouping}.}
\label{tab:measurement_grouping}
\begin{ruledtabular}
\begin{tabular}{lrrrrr}
 & & & \multicolumn{3}{c}{Reduction factor $R_{\mathrm{set}}=M/G$} \\
\cline{4-6}
Model space & $N_q$ & $M$ & QWC & FC greedy & Majorana greedy \\
\hline
$p$ (CK) & 12 & 974 & 3.7 & 20.7 & 20.3 \\
$sd$ (USDB) & 24 & 12,868 & 3.6 & 40.7 & 33.8 \\
$psd$ (YSOX) & 36 & 36,562 & 2.9 & 62.1 & 47.7 \\
$pf$ (GXPF1A) & 40 & 82,860 & 3.9 & 72.9 & 58.0 \\
$sd$--$pf$ (SDPF-MU) & 64 & 299,024 & 2.9 & 116.9 & 88.3 \\
$sd$--$pf$--$sdg$ (SDPFSDG-MU) & 124 & 3,763,622 & --- & 239.6 & 212.2 \\
NN, $e_{\max}=0$ & 4 & 18 & 2.0 & 9.0 & 6.0 \\
NN, $e_{\max}=1$ & 16 & 1,792 & 2.9 & 32.0 & 25.6 \\
NN, $e_{\max}=2$ & 40 & 57,284 & 3.4 & 76.6 & 55.9 \\
NN, $e_{\max}=3$ & 80 & 766,680 & 3.7 & 146.1 & 127.1 \\
NN, $e_{\max}=4$ & 140 & 6,180,598 & --- & 236.0 & 268.4 \\
NN, $e_{\max}=5$ & 224 & 35,470,272 & --- & --- & 515.1 \\
NN+3N, $e_{\max}=1$ & 16 & 51,376 & 4.3 & 51.2 & 28.4 \\
NN+3N, $e_{\max}=2$ & 40 & 15,343,844 & --- & --- & 201.0 \\
\end{tabular}
\end{ruledtabular}
\end{table*}

Higher-order and randomized product formulas, which are not included in
our estimates, are discussed in Sec.~\ref{subsec:extensions}.
Appendix~\ref{app:higher_order} tests first-, second-, and fourth-order
formulas for a six-qubit Hamiltonian.

\subsection{Grouping of encoded Hamiltonian terms}
\label{sec:measurement_grouping}
For Hamiltonian measurements, we separate the identity contribution and write
$H=h_I I+\sum_{\alpha=1}^{M}h_\alpha P_\alpha$.
Let $G$ denote the number of groups of simultaneously measurable nonidentity
Pauli terms. We define the setting reduction factor as
\begin{align}
 R_{\mathrm{set}}=M/G,\qquad N_{\mathrm{circ}}=G.
 \label{eq:Nred}
\end{align}
Qubit-wise commuting (QWC) groups share a single-qubit Pauli measurement axis
on every qubit. Fully commuting (FC) groups allow entangling Clifford basis
changes. Majorana pairing provides a structured FC construction: disjoint
pairs $i\gamma_a\gamma_b$ commute, and products of these pairs are read from
the same measurement outcome~\cite{BonetMonroig2020Measurements}.
These basis changes require Clifford gates; their two-qubit gate cost is
not counted as a T-gate cost in the present model.

Table~\ref{tab:measurement_grouping} benchmarks coefficient-sorted QWC,
FC, and Majorana-pairing greedy constructions on the same retained terms
of each nuclear Hamiltonian. The retained count $M$ excludes identity
and applies a $10^{-8}$ MeV coefficient cutoff, so it differs from the
structural count in Table~\ref{tab:ModelSpace}. The NN and NN+3N inputs
use $\lambda_{\rm SRG}=2.0\,\mathrm{fm}^{-1}$ and $\hbar\omega=20$ MeV,
with $E_{3\max}=3e_{\max}$ for NN+3N.
The QWC reductions are typically about three
to four, whereas the FC and Majorana reductions increase substantially
with model-space size. For example, the FC reduction grows from $20.7$
in the $p$ shell to $239.6$ in the $sd$--$pf$--$sdg$ space. With NN+3N
interactions, Majorana greedy gives reductions of $28.4$ and $201.0$ at
$e_{\max}=1$ and $2$, respectively. These are achieved groupings, not
proofs of minimum setting counts. Setting counts also do not determine
the sampling cost at a fixed precision, which depends on the group
variances and covariances~\cite{Yen2022Measurements}.

\begin{figure*}
  \centering
   \begin{minipage}{0.48\textwidth}
  \includegraphics[width=\linewidth]{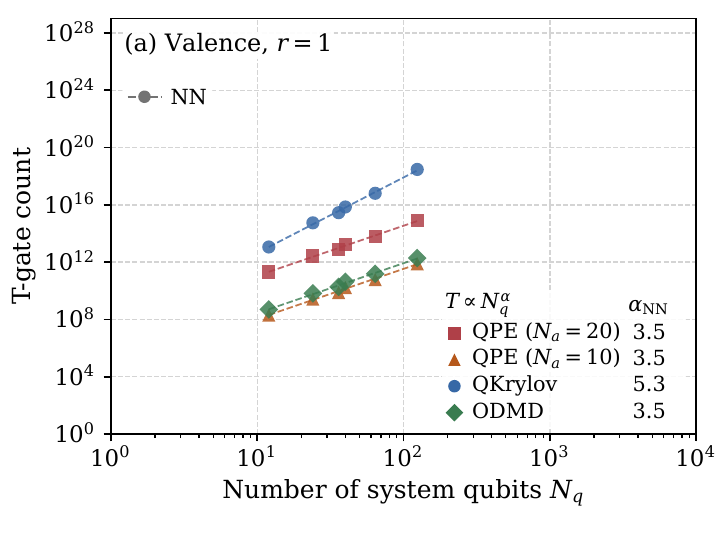}
  \end{minipage}\hfill
   \begin{minipage}{0.48\textwidth}
  \includegraphics[width=\linewidth]{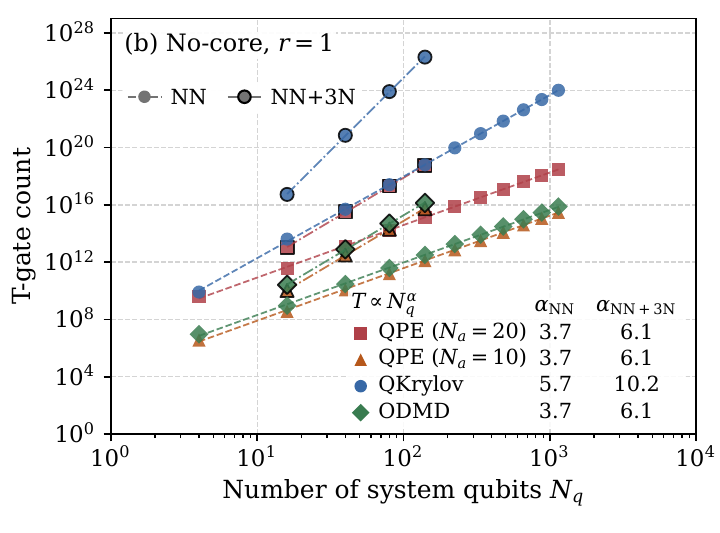}
  \end{minipage}
  \caption{Logical T-gate count $T$ at fixed algorithm parameters versus
  the number of system qubits $N_q$ for (a) valence spaces and (b) no-core
  spaces. The rightmost valence points use the full $sd$--$pf$--$sdg$ space
  ($124$ qubits). Each basic interval uses one first-order Trotter step
  ($r=1$), with $100$ T gates per rotation and one shot per circuit. QPE
  uses $N_a=10$ or $20$ phase qubits.
  For QKrylov, we use 50 vectors and count measurement settings
  with the coefficient-independent Majorana-pairing bound of Eq.~\eqref{eq:grouping_cost_prescription}.
  Table~\ref{tab:measurement_grouping} reports the reduction factors $M/G$ achieved by greedy grouping, where $M$ is the number of retained nonidentity Pauli terms and $G$ is the number of measurement settings.
  Lines show
  fits of $T\propto N_q^\alpha$ over the sampled range for each series.
  For each method, the legend gives $\alpha_{\mathrm{NN}}$ for NN alone
  and, in (b), $\alpha_{\mathrm{NN+3N}}$ when 3N terms are included.
  Nuclear state preparation is excluded.
  \label{fig:Qresource_summary}}
\end{figure*}

For Fig.~\ref{fig:Qresource_summary}(a,b), we select one constructive
Majorana-pairing prescription that applies to every plotted space without
a search over the full Hamiltonian. Its setting upper bounds
$B_2(N_q)$ for NN and $B_3(N_q)$ for NN+3N are derived in
Appendix~\ref{app:grouping_prescription}. We use
\begin{align}
 N_{\mathrm{circ}}=\min\{M_{\mathrm{str}},B_k(N_q)\},
 \label{eq:grouping_cost_prescription}
\end{align}
where $M_{\mathrm{str}}=N_{\hat H}-1$ is the nonidentity structural count
from Table~\ref{tab:ModelSpace}, and $k=2$ or $3$.
The minimum includes measuring each term separately as a fallback.
This prescription retains empty or duplicate settings in the cover bound
and can therefore be looser than the Hamiltonian-specific greedy results.
It supplies a conservative setting count at the large spaces where greedy
grouping has not been completed, rather than extrapolating a constant
empirical reduction factor.

For a diagonal QKrylov matrix element, one prepares a Krylov vector and
measures these $G$ settings. A complex off-diagonal element uses the same
system settings with ancilla $X$ and $Y$ readout, giving $2G$ circuits.
This factor of two is already included in the QKrylov formula in
Table~\ref{tab:Qresources_Methods}. The identity contribution is
$h_I\tilde N_{mn}$; the ancilla marginal of a Hamiltonian measurement can
also supply the overlap $\tilde N_{mn}$. Measurement grouping changes
the QKrylov circuit count, but does not enter the QPE formulas in that table.

\subsection{Cost formulas for major algorithms}
\label{sec:cost_formulas}

\begin{table}
\centering
\caption{Logical T-gate count formulas for the eigenvalue algorithms.
Parameter definitions and the cost models of Figs.~\ref{fig:Qresource_summary}
and \ref{fig:QPE_fixed_accuracy} are summarized in Table~\ref{tab:cost_model}; derivations are in
Appendices~\ref{app:Tformulas} and \ref{app:qubitized_QPE}.
\label{tab:Qresources_Methods}}
\begin{ruledtabular}
  \begin{tabular}{lc}
  \textbf{Method}  & \textbf{T-gate count} \\ 
  \hline
  QPE (Trotter) &  $N_\mathrm{shot}\left[T_{c\hat{U}} (2^{N_a} - 1) + T_{\mathrm{QFT}^\dagger}\right]$ \\
  QPE (Qubitization) & $N_\mathrm{shot} \left[ T_{cW} (2^{N_a} - 1) + T_{\rm PREP} + T_{\mathrm{QFT}^\dagger}\right] $ \\ 
  QKrylov & $N_\mathrm{shot} T_{\hat{U}} N_{\mathrm{circ}} \,(4 N^3_{\mathrm{iter}}+ N^2_{\mathrm{iter}} - 3N_{\mathrm{iter}})/2$ \\
  ODMD & $N_\mathrm{shot} \, T_{c\hat{U}} \cdot N_{\mathrm{snap}}(N_{\mathrm{snap}} + 1)$ 
  \end{tabular}
\end{ruledtabular}
\end{table}

In Table~\ref{tab:Qresources_Methods}, $N_{\mathrm{shot}}$ is the
number of repetitions of each distinct circuit. QPE uses one circuit
with $N_a$ phase qubits. ODMD uses two circuits for each of its
$N_{\mathrm{snap}}$ nonzero sampling times. For QKrylov, each of the
$N_{\mathrm{circ}}$ measurement settings is used for the
$N_{\mathrm{iter}}$ diagonal elements and for both quadratures of
each distinct off-diagonal pair. One pass therefore takes one shot
for QPE, $2N_{\mathrm{snap}}$ shots for ODMD, and
$N_{\mathrm{circ}}N_{\mathrm{iter}}^2$ shots for QKrylov.
The T-gate count also accounts for the number of evolution intervals
in each circuit.

For Trotterized evolution, $T_{\hat U}$ and
$T_{c\hat U}=2T_{\hat U}$ are the costs of one complete basic
interval, including all its Trotter substeps. For qubitized QPE,
$T_{cW}$ is the cost of one controlled walk and $T_{\rm PREP}$
initializes its ancillas once per spectral sample. Nuclear state
preparation is excluded. $T_{\mathrm{QFT}^\dagger}$ denotes the
inverse-QFT cost for the phase register of the corresponding circuit.

\section{Resource estimates}
\label{sec:results}

\begin{figure*}
  \centering
   \begin{minipage}{0.333\textwidth}
  \includegraphics[width=\linewidth]{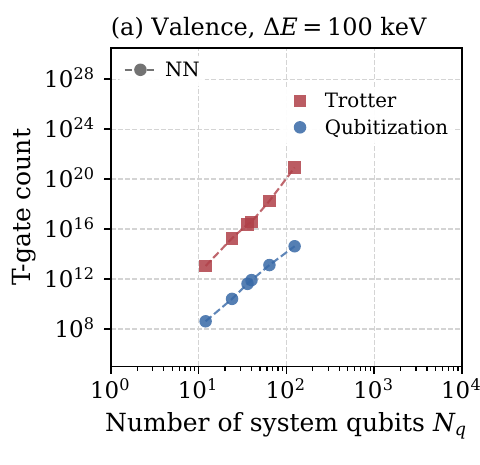}
  \end{minipage}\hfill
   \begin{minipage}{0.333\textwidth}
  \includegraphics[width=\linewidth]{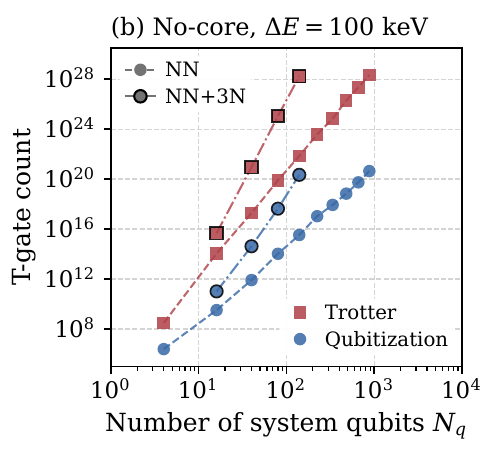}
  \end{minipage}\hfill
   \begin{minipage}{0.333\textwidth}
  \includegraphics[width=\linewidth]{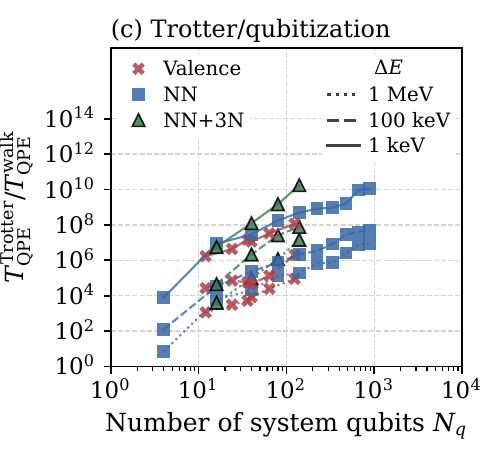}
  \end{minipage}
  \caption{QPE at a fixed absolute energy accuracy relative to each
  retained Hamiltonian. (a,b)~Logical T-gate count of one spectral sample
  at $\Delta E=100$~keV for first-order Trotterization, with step counts
  from the pairwise commutator bound, and for qubitization, in (a) valence
  and (b) no-core spaces. (c)~Ratio of the two T-gate counts at
  $\Delta E=1$~MeV, $100$~keV, and $1$~keV for the valence, no-core NN,
  and no-core NN+3N Hamiltonians. Lines connect the evaluated points and
  are not extrapolated fits. Black marker outlines identify the NN+3N
  cases. The cost model is specified in Appendix~\ref{app:qubitized_QPE};
  nuclear state preparation is excluded.
  \label{fig:QPE_fixed_accuracy}}
\end{figure*}

We now apply the cost model of Sec.~\ref{sec:resource_estimation} to
the benchmark Hamiltonians. Figures~\ref{fig:Qresource_summary} and \ref{fig:QPE_fixed_accuracy}
summarize the logical T-gate counts at fixed algorithm parameters and
for QPE at fixed energy accuracy, respectively.

\subsection{Scaling at fixed algorithm parameters}
\label{sec:Qresource_summary}

We first examine how the logical T-gate count grows with the size of the
nuclear model space. Figure~\ref{fig:Qresource_summary}(a,b) compares
Trotterized QPE, QKrylov, and ODMD as the number of system qubits
$N_q$ increases, with the algorithm parameters held fixed for each
series. Each basic evolution interval consists of one first-order
Trotter step ($r=1$).
Table~\ref{tab:ModelSpace} gives
the structural count $N_{\hat H}$ including the identity, so the step
contains $M_{\mathrm{str}}=N_{\hat H}-1$ Pauli rotations. We use
$T_\epsilon=100$, giving $T_{\hat U}=M_{\mathrm{str}}T_\epsilon$
and $T_{c\hat U}=2T_{\hat U}$.

We use $N_a=10,20$ for QPE;
$N_{\mathrm{iter}}=50$ is the QKrylov dimension and
$N_{\mathrm{snap}}=50$ is the ODMD snapshot count.
For QKrylov, $N_{\mathrm{circ}}$ is the Majorana-pairing setting bound
in Eq.~\eqref{eq:grouping_cost_prescription}, before the ancilla $X/Y$
factor already included in the cost formula.
$N_{\mathrm{shot}}=1$ denotes one repetition of each circuit in the Figure;
the formulas leave this repetition factor explicit.
These two panels do not impose a common final energy-error budget.

Figure~\ref{fig:Qresource_summary}(a,b) shows how the circuit costs of
Trotterized QPE, QKrylov, and ODMD grow with model-space size when the
algorithm parameters and the cost per rotation are held fixed. The
structural term counts and the bound on QKrylov measurement settings
give conservative counts for these circuits.
The horizontal axis counts system qubits, one per single-particle state in the Jordan--Wigner encoding.
At fixed $N_q$, including three-body interactions can substantially increase the cost of an
evolution step. For example, at $e_{\max}=2$ both no-core Hamiltonians use 40 system qubits.
The structural count of nonidentity Pauli terms increases from 57,444 for NN to 15,346,372
for NN+3N (Table~\ref{tab:ModelSpace}), increasing the T-gate count of one first-order Trotter step in panel~(b) by a factor of about 267.
The fitted exponents in panels~(a,b) describe the dependence on $N_q$ over the model spaces sampled.

For QPE and ODMD, the plotted T-gate count is the cost $T_{\hat U}$ of
one Trotter step multiplied by a factor fixed by the algorithm
parameters: $2(2^{N_a}-1)$ for QPE, plus the small inverse-QFT cost, and
$2N_{\mathrm{snap}}(N_{\mathrm{snap}}+1)$ for ODMD. These curves
therefore share the same fitted exponent, and their vertical offsets
reflect only the chosen parameters. For example, one pass through the
50-snapshot ODMD circuits costs about 2.5 times one 10-bit QPE sample
and far less than one 20-bit QPE sample. QKrylov depends on $N_q$ also
through the number of measurement settings $N_{\rm circ}$ in
Eq.~\eqref{eq:grouping_cost_prescription}, which accounts for its larger
exponent. Panels (a,b) thus show how the cost of time evolution grows
with the model space and with 3N forces. They do not rank the algorithms
at a common accuracy, which also requires the shot counts discussed in
Sec.~\ref{sec:finite_shot}.

\subsection{QPE at fixed energy accuracy}
\label{sec:qpe_fixed_accuracy}

We next fix the target energy accuracy and compare the logical
T-gate counts of Trotterized and qubitized QPE
(Fig.~\ref{fig:QPE_fixed_accuracy}). For each benchmark Hamiltonian, both implementations target a total absolute
energy error of $\Delta E=100$~keV relative to its exact eigenenergies.

Resolving two levels separated by less than this tolerance can require a finer energy resolution.
We allow $50$ keV for the uncertainty in reading an energy from the
finite phase register. The other $50$ keV bounds the shift in the energy
itself due to the simulation method. For Trotterized QPE we choose the
number of first-order Trotter substeps to keep the spectral error within
this bound. For qubitized QPE we choose the coefficient precision so
that the sum of the absolute rounding errors over all Pauli
coefficients is at most $50$ keV.

For each retained Hamiltonian, the inputs are the nonidentity term
count $M$, the coefficient norm $\lambda'=\sum_{j\ne I}|h_j|$,
and the pairwise commutator coefficient $C_{\rm pair}$
(Table~\ref{tab:Trotter_Error}). We restore
the identity shift classically. For Trotterized QPE we set
$\delta t=\pi/(2\lambda')$ and use $C_{\rm pair}$ to choose the
number of substeps per interval. The phase-register size $N_a$ in
Table~\ref{tab:Qresources_Methods} is chosen separately for the two
implementations: it is $m_T$ for Trotterized QPE and $m_W$ for
qubitized QPE, as given in Eq.~\eqref{eq:qpe_common_bits}.

\begin{table*}
\caption{Finite-shot resources per ground-state energy estimate for $^{6}$He and $^{22}$O using ideal circuits.
Size counts phase qubits for QPE, nonzero sampling times for ODMD, and basis states for QKrylov.
For each nucleus and method, Shots/setting gives the number of repetitions
chosen for each circuit before validation.
The number of circuit executions per estimate (Total shots) equals the number of
distinct circuits (Settings) times the shots per setting.
For QPE we return the sample lower quartile, chosen because the 25th percentile of
each calibrated readout distribution lies within the target energy window.
QKrylov uses greedy Majorana pairing.
Each interval $\delta t$ comprises $r=4$ first-order Trotter substeps.
Max.\ substeps counts sequential substeps in the longest circuit.
$W$ counts time-evolution Pauli rotations over all shots, with weight one for uncontrolled rotations
and two for controlled rotations. The ratio $W/W_{\rm QPE}$ uses QPE for the same nucleus as its reference.
Success gives the number of estimates within $100$~keV of the exact ground-state energy in 1,200 independent trials per row
(target success probability $95\%$).}
\label{tab:r21_fixed_sampling}
\begin{ruledtabular}
\begin{tabular}{llrrrrrrrr}
Nucleus & Method & Size & $\delta t$ ($\mathrm{MeV}^{-1}$) & Settings & Shots/setting & Total shots & Max.\ substeps & $W/W_{\rm QPE}$ & Success \\
\hline
$^{6}$He & QPE     &  9 & 0.250   &     1 &     56 &          56 & 2,044 & 1                  & 1,185 \\
         & ODMD    & 32 & 0.250   &    64 &     64 &       4,096 &   128 & 2.36               & 1,198 \\
         & QKrylov &  4 & 0.375   &   256 & 16,384 &   4,194,304 &    12 & 247                & 1,200 \\
$^{22}$O & QPE     & 11 & 0.07409 &     1 &      7 &           7 & 8,188 & 1                  & 1,184 \\
         & ODMD    & 64 & 0.055   &   128 &     16 &       2,048 &   256 & 4.65               & 1,181 \\
         & QKrylov &  7 & 0.180   & 5,586 & 32,768 & 183,042,048 &    24 & $4.11\times10^4$   & 1,200 \\
\end{tabular}
\end{ruledtabular}
\end{table*}

Approximating single-qubit rotations with Clifford and T gates also
changes the QPE output probabilities. For each complete QPE circuit,
we bound the operator-norm difference between its unitaries before
and after rotation synthesis by $10^{-3}$. For a circuit with $n_R$
synthesized rotations, we allow an error of $10^{-3}/n_R$ per rotation.
Circuits with more rotations therefore require higher precision and
generally more T gates per rotation. The counts are per spectral sample.
Given an eigenstate of the encoded operator, both methods return an
energy within the specified tolerance with probability above $0.80$
in this model. We assume equal usable target-state overlap and exclude
nuclear state preparation.

Figure~\ref{fig:QPE_fixed_accuracy}(a,b) shows the T-gate count of one
QPE spectral sample at $\Delta E=100$~keV. These calculations use the
coefficient-screened Hamiltonians of Table~\ref{tab:Trotter_Error},
whose term counts can differ from the structural counts used in
Fig.~\ref{fig:Qresource_summary}. Panel (a) covers the same six valence
spaces as Fig.~\ref{fig:Qresource_summary}(a), from the $p$ shell
($12$ qubits) to the $sd$--$pf$--$sdg$ space ($124$ qubits). Panel (b)
covers no-core NN spaces through $e_{\max}=9$ ($880$ qubits) and NN+3N
spaces through $e_{\max}=4$ ($140$ qubits). Black marker outlines and
dash-dotted lines identify the NN+3N cases.

The leading dependence of these costs on the Hamiltonian follows from
the cost model of Appendix~\ref{app:qubitized_QPE}. With
$\delta t=\pi/(2\lambda')$, the number of Trotter substeps per interval
scales as $r\propto C_{\rm pair}/(\lambda'\Delta E)$ and the number of
controlled intervals as $2^{m_T}\propto\lambda'/\Delta E$, so that
$T^{T}_{\rm QPE}\propto M C_{\rm pair}/\Delta E^{2}$. The number of walk
calls scales as $2^{m_W}\propto\lambda'/\Delta E$ and the cost of one
walk as $M$, so that $T^{W}_{\rm QPE}\propto M\lambda'/\Delta E$. Their
ratio therefore scales as $C_{\rm pair}/(\lambda'\Delta E)$, up to
logarithmic factors from rotation synthesis and the rounding of register
sizes. Fits over the evaluated cases give $T^{T}_{\rm QPE}\propto
N_q^{7.7}$, $N_q^{8.2}$, and $N_q^{13.4}$ for the valence, no-core NN
($e_{\max}\geq1$), and no-core NN+3N series, compared with
$N_q^{6.0}$, $N_q^{6.4}$, and $N_q^{9.9}$ for qubitized QPE.

At $\Delta E=100$~keV, the estimated T-gate count per spectral sample is
lower for qubitized QPE in all 20 Hamiltonians evaluated. The ratio
of the first-order Trotter estimate to the qubitization estimate
ranges from about 126 for the no-core NN Hamiltonian at $e_{\max}=0$
to $7.8\times10^7$ for NN+3N at $e_{\max}=4$.
At $e_{\max}=2$, where both no-core Hamiltonians act on 40 qubits,
adding the 3N interaction increases the fixed-accuracy T-gate count by a
factor of about $4.7\times10^{3}$ for Trotterized QPE and
$5.1\times10^{2}$ for qubitized QPE, compared with the factor of 267 in
the cost of one Trotter step at fixed parameters. The larger Trotter
factor reflects the simultaneous growth of $M$ and $C_{\rm pair}$.

Figure~\ref{fig:QPE_fixed_accuracy}(c) shows the ratios at total energy errors
of $1$~MeV, $100$~keV, and $1$~keV. Lines connect evaluated cases within
each interaction family; they are not extrapolated fits.
For these benchmarks and target accuracies, the effective gate-cost
factor $2M T_R^T/T_{cW}$ of Appendix~\ref{app:qubitized_QPE} ranges from
about 4.8 to 48.3. At a total energy error of 1 keV, the calculated
ratios range from $7.9\times10^3$ to $1.8\times10^{10}$.
No crossing of unity occurs among the evaluated cases at any of the
three accuracies. For the $p$-shell CKpot Hamiltonian, tightening
$\Delta E$ from $1$~MeV to $100$~keV increases the estimated T-gate
count per sample by a factor of about $190$ for first-order Trotterized
QPE and $8$ for qubitized QPE, consistent with the $\Delta E^{-2}$ and
$\Delta E^{-1}$ dependences above up to the rounding of register sizes
and synthesis costs. These comparisons use first-order Trotterization
with step counts from the pairwise bound and a specified walk
implementation; Sec.~\ref{subsec:limitations} summarizes why they do not
establish the relative costs of optimized implementations.

\subsection{Finite-shot ground-state energy estimates}
\label{sec:finite_shot}

We then asked how many circuit executions QPE, QKrylov, and ODMD
require to obtain one ground-state energy estimate. For both
$^{6}$He and $^{22}$O, the target was an estimate within $100$~keV
of the exact ground-state energy with a success probability of at
least $95\%$. For each nucleus we use the Jordan--Wigner encoding of
the neutron valence Hamiltonian and obtain the reference energy by exact
diagonalization in the complete $M_J=0$ sector at fixed neutron number.
Within each nucleus we used the same initial determinant and four
first-order Trotter substeps per basic interval. For QPE, the determinant
also populates excited states. We used the known ground-state energies
to choose a low percentile that selects the ground-state peak in both
nuclei and to set the number of readouts per estimate
(Appendix~\ref{app:r21_fixed_sampling}). For ODMD and QKrylov we chose
a uniform number of shots per circuit for each method and nucleus.
With these choices fixed, we tested how often each method returned an
energy in the target window on independent trials.

Table~\ref{tab:r21_fixed_sampling} gives both the number of circuit executions
per estimate and the maximum number of sequential Trotter substeps in one
circuit at the validated settings. For
$^{22}$O, QPE combined seven readouts of a circuit with 8,188 substeps.
QKrylov limited each circuit to at most 24 substeps, but repeated
measurements across 5,586 circuits to obtain one estimate. In both nuclei
QPE used the fewest shots and QKrylov the shortest coherent evolution
among the protocols tested.

These shot counts change the comparison suggested by
Fig.~\ref{fig:Qresource_summary}(a,b), which counts one shot per
circuit and therefore compares only the cost of one pass through each
method's circuits. At the validated budgets for ${}^{22}$O, the total
time-evolution work of ODMD and QKrylov is $4.65$ and
$4.11\times10^{4}$ times that of QPE, respectively, while their longest
circuits are 32 and 341 times shorter. Thus the number of shots needed
to reach a common accuracy, rather than the cost of one pass, determines
the ordering of the total work in these examples. The relative costs
can change with the spectrum, the ground-state weight of the input
state, the estimator, and the allocation of shots across circuits.

\subsection{State-preparation overhead}
\label{sec:State_prep}

State preparation is a central bottleneck for early fault-tolerant quantum simulations of many-body Hamiltonians.
In practice, scalable ground-state preparation algorithms rely on two key assumptions: (i) the availability of a trial state with nonzero overlap with the target ground state, and (ii) coarse spectral information within the energy window of interest, namely estimates of the lowest and highest eigenvalues $(E_{\min}, E_{\max})$ and the spectral gap $\Delta$ above the ground state. Given this information, it is possible to design spectral filters implemented via quantum signal processing \cite{Dong:2022} or related constructions \cite{Choi:2021,Stetcu:2023} that approximate a ground-state projector, suppressing excited-state components while amplifying the ground-state overlap. This framework encompasses a broad class of filtering-based methods, distinct from variational approaches, which lack exactness guarantees, and adiabatic evolution, which can require prohibitively long coherent evolution times when gaps are small.

The resource estimates in Figs.~\ref{fig:Qresource_summary} and
\ref{fig:QPE_fixed_accuracy} assume that the input state is already
prepared (Sec.~\ref{sec:assumptions}). The finite-shot calculations
(Sec.~\ref{sec:finite_shot}) and statevector demonstrations
(Sec.~\ref{sec:demonstrations}) use single Slater determinants.
For ${}^{22}$O, both use the filled $0d_{5/2}$ neutron subshell.

The first direction is energy filtering. Among filtering methods, the quantum projection filter (QPF) of
Ref.~\cite{Stetcu:2023} combines symmetry-sector projection with energy
filtering. By removing states outside the target symmetry sector, the
method can increase the relevant spectral gap and thereby reduce the
cost of the subsequent energy filter. The required evolution time
scales as $1/\Delta_{\mathrm{eff}}$, where $\Delta_{\mathrm{eff}}$ is
the gap within the projected symmetry sector. The dependence on the
initial overlap $\alpha$ between the trial state and the exact ground
state leads to an expected repetition cost that scales as
$\mathcal{O}(1/\alpha^2)$. These features make QPF a candidate for
early fault-tolerant regimes, where symmetry information is often
available and circuit depth, rather than qubit count, is the dominant
constraint.

Appendix~\ref{app:qpf} summarizes the block-encoding construction of the
filter and the resulting ground-state overlap.
Reference~\cite{Stetcu:2023} optimized the filter at a total evolution
time $\sum_i t_i=\pi/\Delta_{\rm eff}$. With this choice, halving the assumed
gap doubles the time budget. Overestimating the gap can leave low-lying
excited states in the filtered state. The achieved fidelity also depends on
the individual times, phases, and initial spectral weights in
Eq.~\eqref{eq:qpf_overlap}.
Figure~\ref{fig:QPE_fixed_accuracy} excludes nuclear state preparation, so its
QPE estimates require no assumption about this gap.

A second direction, not pursued here, starts from a classical
approximation of the ground state. As a representative
preparation-cost estimate, the recent preprint
by Gibbs et al.~\cite{Gibbs:2026} combines classical DMRG approximations
as matrix product states (MPS) with variational circuit compilation
and Clifford$+$T synthesis. For ${}^{142,143}$Ce in a 76-qubit valence
space with the KHHE interaction, circuits with approximately
$2\times10^4$ T gates reach an estimated lower bound on the
ground-state overlap amplitude of $\alpha\sim0.5$.
This bound uses an extrapolation of the MPS overlap to infinite bond
dimension, rather than exact diagonalization of these large systems.
The T-gate count is for one preparation; classical DMRG and circuit
optimization are separate costs. These results provide a concrete
valence-space example, not a preparation-cost bound for the no-core
NN and NN+3N spaces studied here.

For independent QPE runs without filtering or amplitude amplification,
the cost per run is $T_{\rm prep}+T_{\rm QPE}$, where $T_{\rm prep}$
denotes nuclear state preparation and $T_{\rm QPE}$ the subsequent
phase-estimation circuit. If $q$ is the conditional probability of
obtaining a target-energy sample from the target eigenstate, the
probability of a successful sample from the target component per run is
$\alpha^2q$. Using $\alpha=0.5$ for illustration, the overlap
alone therefore gives an expected waiting count of four runs
when $q=1$; finite phase-estimation success increases this count.
Preparation is paid on every run. This example does not establish
that preparation and repetition costs are negligible in our benchmarks.
For QKrylov and ODMD, preparation is likewise required on each measurement
circuit execution, but their statistical overhead cannot be inferred
from this QPE repetition factor.

\begin{figure}
  \centering
\includegraphics[width=0.98\linewidth]{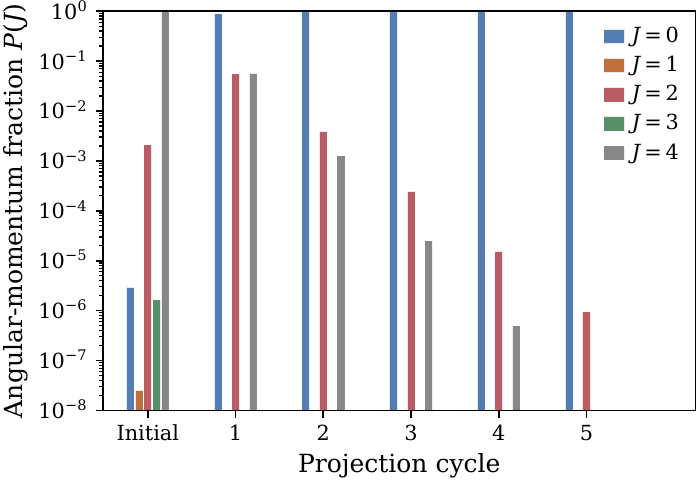}
 \caption{
Angular-momentum composition of the ${}^{18}$O trial state before
projection and after each of five projection cycles. Each cycle
consists of a $J_z=0$ projection followed by a $J_x=0$ projection.
The bars show the weight $P(J)$ of each total-angular-momentum
sector in the normalized state at that stage. For the projected
states, these weights are conditional on successful postselection
through the indicated cycle.
\label{fig:state_prep_J0}
 }
\end{figure}

A third direction is the angular-momentum projection method of
Ref.~\cite{Rule:2024}, which prepares a trial state in the $J=0$
sector. Its circuit uses the one-body operators $J_z$ and $J_x$
to project alternately onto zero angular-momentum projection
along the two axes. A $J=0$ component is preserved by both
projections, while repeated cycles suppress components with $J>0$.
Because the projection selects a symmetry sector rather than an
energy, energy filters such as QPF could further distinguish
eigenstates within the selected sector. We do not use this projection
in the resource estimates or in the demonstrations of
Sec.~\ref{sec:demonstrations}; the example below only illustrates
its postselection overhead.

Figure~\ref{fig:state_prep_J0} shows a noise-free statevector
simulation of the projection for ${}^{18}$O, with two valence neutrons
in the $sd$ shell above an inert ${}^{16}$O core. As the input of this
illustration, we apply particle-number-conserving Givens rotations with
random angles to a filling configuration. This state is dominated by
$J=4$ and has only a small $J=0$ component, $P(J=0)=2.97\times10^{-6}$.

After five cycles, the normalized postselected state has
$P(J=0)>0.99999$. For this initial state, the cumulative
probability of obtaining all required ancilla outcomes is
$2.97\times10^{-6}$. Comparison with exact diagonalization of
the USDB Hamiltonian~\cite{USDB} gives a squared ground-state
overlap of $0.776$ for the postselected state. Two excited
$J=0$ states carry nearly all of the remaining weight.
For an exact projection, the cumulative success probability equals the
$J=0$ weight of the initial state. Here the initial weight is
$P(J=0)=2.97\times10^{-6}$, so about $3.4\times10^{5}$ preparation
attempts are needed on average. The overhead of postselected projection
is therefore set by the initial state rather than by the projection
circuit; a trial state with a larger $J=0$ weight reduces it
proportionally.
Whether projection lowers the total cost of an energy estimate would
also depend on how the trial state is prepared and on any subsequent
energy filtering, which we do not assess here.


\section{Demonstrations of the workflow via NuQuLib}
\label{sec:demonstrations}

We use NuQuLib and Qiskit~\cite{Qiskit} to simulate small nuclear
systems with a noise-free statevector simulator.
The NuQuLib tutorial notebook contains the examples in this
section~\cite{Repo_NuQuLib,*Zenodo_NuQuLib}.
Both QPE demonstrations use $N_a=6$ ancilla qubits and direct matrix exponentiation for controlled Hamiltonian evolution.
Their reference spectra are obtained by exact diagonalization in the same
particle-number, $M$, and parity sectors as the simulations.
These small-system simulations isolate finite-ancilla resolution and initial-state overlap from product-formula error. They do not implement the gate-level resource model of Sec.~\ref{sec:resource_estimation}.

This section presents QPE examples for a three-neutron system and
${}^{6}$He, followed by QKrylov and ODMD for ${}^{22}$O. The ${}^{6}$He
Hamiltonian is the six-qubit neutron-space Hamiltonian also used in the
finite-shot estimates of Sec.~\ref{sec:finite_shot} and the
product-formula test of Appendix~\ref{app:higher_order}; here the
controlled evolution is evaluated exactly.
The angular-momentum projection example for ${}^{18}$O illustrates a
possible state-preparation direction and is presented in
Sec.~\ref{sec:State_prep}.

\subsection{Quantum Phase Estimation (QPE)}

\begin{figure}[htbp]
\centering
\includegraphics[width=0.48\textwidth]{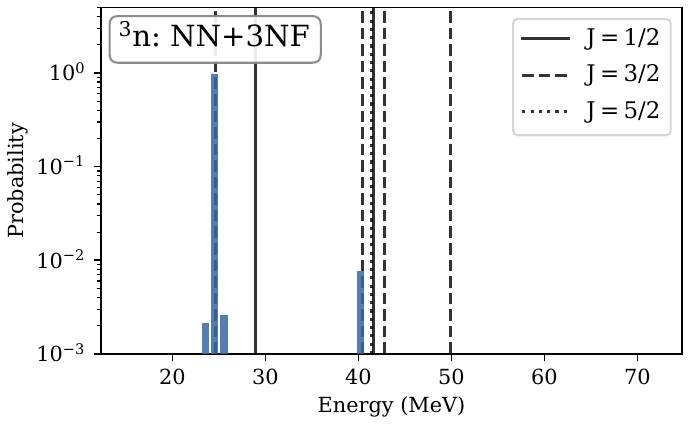}
\caption{\label{fig:QPE_3n_e1}
QPE results for the three-neutron system with NN + 3N interactions with $e_{\max}=1$ and $E_{3\max}=3$.
We use $N_a=6$ ancilla qubits and a basic evolution interval
$\delta t=-0.1\,\mathrm{MeV}^{-1}$.
Controlled evolution is evaluated by direct matrix exponentiation.
Bars give the readout probabilities. Vertical lines mark the exact eigenenergies
in the $M=1/2$, negative-parity sector, with line styles indicating $J$.
}
\end{figure}

\begin{figure}[htbp]
\centering
\includegraphics[width=0.48\textwidth]{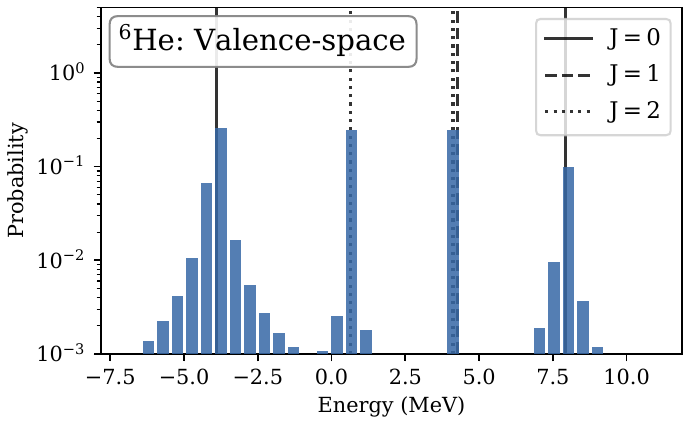}
\caption{\label{fig:QPE_6He}
QPE readout probabilities for ${}^{6}$He in the valence $p$ shell with the CKpot~\cite{CohenKurath} interaction. Vertical lines mark the exact eigenenergies in the $M=0$, positive-parity sector.
We use $N_a=6$ ancilla qubits and $\delta t=0.2\,\mathrm{MeV}^{-1}$, with controlled evolution evaluated by direct matrix exponentiation.
}
\end{figure}

In Fig.~\ref{fig:QPE_3n_e1}, we show the QPE results for three-neutron system
in a model space with $e_{\max}=1$ and $E_{3\max}=3$,
retaining the full three-body space within this single-particle truncation
and imposing no additional $N_\mathrm{max}$ truncation.
The adopted Hamiltonian consists of the EM500 NN~\cite{EMPRC, EMrev1} and 3N (lnl)~\cite{Soma_LNL} interactions softened
via similarity renormalization group (SRG) with $\lambda_\mathrm{SRG} =2.0\ \mathrm{fm}^{-1}$.
The harmonic oscillator frequency is set to $\hbar \omega = 20$ MeV and we did not separate
the center-of-mass motion in this demonstration.
The negative interval $\delta t=-0.1\,\mathrm{MeV}^{-1}$ maps the
positive energies shown for this unbound system directly to phases
in $[0,1)$ under $U=\exp(-iH\delta t)$; energies are recovered as
$E=-2\pi\phi/\delta t$.
The initial configuration has two neutrons filling the $0s_{1/2}$
subshell and a third neutron in the $0p_{3/2}$ orbital with
$m_j=1/2$. The closed $0s_{1/2}$ pair carries $J=0$, so this
configuration has $J=3/2$ and negative parity.
The vertical lines show all exact eigenenergies in the
$M=1/2$, negative-parity sector. Only $J=3/2$ eigenstates
contribute to the QPE distribution. The dominant peak at
$24.54$~MeV is the readout bin nearest the lowest $J=3/2$
eigenenergy, $24.59$~MeV.

The valence-space counterpart in Fig.~\ref{fig:QPE_6He} is ${}^{6}$He
with the CKpot interaction~\cite{CohenKurath}.
The problem contains two valence neutrons in the $p$ shell above an inert
${}^{4}$He core, with $M=0$ and positive parity.
For this example we evolve $H-sI$ with $s=10\,\mathrm{MeV}$.
All five eigenenergies in this sector lie below $s$, and
$\delta t=0.2\,\mathrm{MeV}^{-1}$ maps them into the phase
interval $[0,1)$ without wrapping. The energies plotted in
Fig.~\ref{fig:QPE_6He} are recovered from the measured phase
$\widetilde{\phi}$ as $E=s-2\pi\widetilde{\phi}/\delta t$.

The initial state occupies the time-reversed $m_j=\pm 3/2$
orbitals of the $0p_{3/2}$ subshell. Two neutrons in this
subshell can couple only to $J=0$ or $2$. The initial state
has nonzero overlap with both $J=0$ and both $J=2$ eigenstates,
which account for the four peaks in Fig.~\ref{fig:QPE_6He}.
The fourth eigenstate in order of energy has $J=1$ and therefore
does not contribute to the QPE distribution.

\subsection{Quantum Krylov Subspace Method and Observable Dynamic Mode Decomposition}

\begin{figure}[htbp]
\centering
\includegraphics[width=\linewidth]{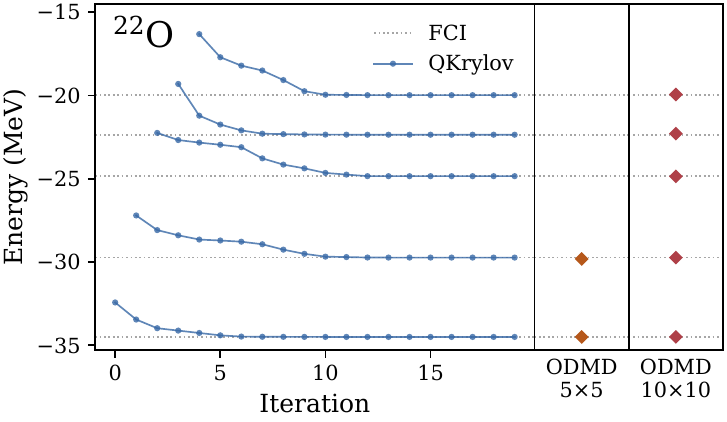}
\caption{\label{fig:QKrylov_ODMD_22O}
Low-lying $J=0$ energies of ${}^{22}$O in the neutron $sd$ shell.
Dotted lines mark the five lowest $J=0$ FCI energies.
Circles show the five lowest QKrylov energies, or all available energies
when the basis contains fewer than five states. Iteration zero contains
only the initial determinant. Diamonds show the two energies retained by
the $5\times5$ ODMD fit and the five lowest of the ten retained by the
$10\times10$ fit.}
\end{figure}

Figure~\ref{fig:QKrylov_ODMD_22O} uses the same 12-qubit
Jordan--Wigner Hamiltonian and initial determinant as the ${}^{22}$O
finite-shot benchmark of Sec.~\ref{sec:finite_shot}. Six valence
neutrons interact through USDB~\cite{USDB} above an inert ${}^{16}$O
core. Full configuration interaction in the 142-dimensional $M=0$,
positive-parity sector gives a ground-state energy of $-34.497871$~MeV.
The filled $0d_{5/2}$ subshell has $J=0$ and a squared ground-state
overlap of $0.75$. Since time evolution preserves $J$, only $J=0$
levels contribute to the signal.

We apply the matrix exponential of the sparse Hamiltonian to the initial
state on the full qubit register. Inner products of the resulting
statevectors give the Krylov overlap and Hamiltonian matrices and the ODMD
autocorrelation. This evaluates ideal evolution and expectation values
directly. The finite-shot benchmark uses first-order Trotter evolution
and samples measurement outcomes.

For QKrylov we use $t_k=k\delta t$ with $\delta t=0.4~\mathrm{MeV}^{-1}$
and $k=0,\ldots,19$. Keeping overlap eigenvalues larger than $10^{-10}$
times the largest leaves 14 independent directions at the final iteration.
The lowest Ritz value agrees with the FCI energy to the six decimal
places quoted above.

For ODMD we evolve $H-sI$ with $s=-20$~MeV and
$\delta t=0.1~\mathrm{MeV}^{-1}$. We form the Hankel matrices of
Eq.~\eqref{eq:Hankel} from the first 10 or 20 autocorrelation values,
including $o_0=1$. We retain singular values above $10^{-10}$ times the
largest and modes satisfying $\bigl||z|-1\bigr|\leq0.05$.
The energies follow from $E=s-\arg(z)/\delta t$. The $5\times5$ fit
gives a ground-state energy of $-34.498861$~MeV, within $1$~keV of
FCI. The $10\times10$ fit gives $-34.497871$~MeV and resolves
additional excited states.

\section{Discussion and Outlook}
\label{sec:outlook}

In this work, we presented a framework for benchmarking quantum algorithms
for nuclear many-body systems, implemented through the NuQuLib package.
The valence-space and no-core Hamiltonians let us examine how the
choice of model space and the inclusion of 3N forces affect circuit
costs. Figure~\ref{fig:Qresource_summary}(a,b) compares Trotterized
QPE, QKrylov, and ODMD at fixed algorithm parameters. For example,
in the 40-qubit no-core space, including 3N forces increases the
estimated T-gate count of one first-order Trotter step by a factor of
about 267. Figure~\ref{fig:QPE_fixed_accuracy} compares two QPE implementations at a common
energy accuracy of $100$~keV relative to each retained Hamiltonian.
For all 20 Hamiltonians tested, qubitized QPE had a lower estimated
T-gate count per spectral sample than first-order Trotterized QPE with
step counts set by the pairwise commutator bound.

The finite-shot examples show why the cost of an individual circuit
gives only part of the cost of a ground-state energy estimate. In
${}^{6}$He and ${}^{22}$O we combined QPE readouts using a percentile
calibrated for the two initial determinants. ODMD and QKrylov instead
combined measurements across times and matrix elements, respectively.
Once the three protocols met the same accuracy and success target,
QPE used the fewest circuit executions per estimate and had the
longest coherent evolution per shot. The relative costs can change
with the spectrum, the input state's ground-state weight, the estimator,
and the allocation of shots across circuits. The statevector
demonstrations in Sec.~\ref{sec:demonstrations} also illustrate how
the choice of trial state affects which eigenstates can be resolved.

\subsection{Scope and limitations}
\label{subsec:limitations}

The resource estimates rest on several simplifications, which we
collect here. (i)~The Hamiltonians are encoded with the Jordan--Wigner
transformation on the full single-particle register, without symmetry
tapering. (ii)~Trotterized costs use the first-order product formula,
and in Fig.~\ref{fig:QPE_fixed_accuracy}
its step count follows the pairwise commutator bound, which sums
commutator norms and discards cancellations. Both choices make the
Trotter estimates conservative. (iii)~Nuclear state preparation is
excluded from Figs.~\ref{fig:Qresource_summary} and
\ref{fig:QPE_fixed_accuracy}; Sec.~\ref{sec:State_prep} gives a literature example and an illustrative projection overhead for ${}^{18}$O.
(iv)~Figure~\ref{fig:Qresource_summary}(a,b) uses one shot per circuit;
the finite-shot calculations of Sec.~\ref{sec:finite_shot} show that
shot counts can change the ordering of the methods. (v)~The logical T-gate counts in
Figs.~\ref{fig:Qresource_summary} and \ref{fig:QPE_fixed_accuracy} exclude
Clifford gates, error correction, and compilation optimizations.
The finite-shot comparison uses shot counts and the weighted number $W$
of time-evolution Pauli rotations before gate synthesis. (vi)~Energy accuracies refer to the
encoded, coefficient-screened Hamiltonian, not to the convergence of
nuclear observables with respect to the interaction or model space.
Consequently, the large Trotter-to-qubitization ratios in
Fig.~\ref{fig:QPE_fixed_accuracy}(c) compare the specified implementations.
They do not establish a crossover between optimized algorithms or the
total cost of estimating a nuclear ground-state energy.

\subsection{Extensions not included in the estimates}
\label{subsec:extensions}

Higher-order product formulas offer a different balance between accuracy
and cost per step. For a formula of order $p$, the global operator-norm
error at fixed evolution time scales as
$\epsilon_p(t,r)=\mathcal{O}(K_p|t|^{p+1}/r^p)$ in the small-step regime,
where $K_p$ depends on nested commutators and the term ordering~\cite{Childs:2021}.
The standard first, second, and fourth-order constructions use a linear number of Pauli rotations per step. This is  $M$, $2M-1$, and $5(2M-1)$ respectively,
before combining rotations across recursive blocks or steps.
Thus, a higher order can reduce the total rotation count needed for a
specified error despite a more expensive step. The T-gate count also depends
on rotation synthesis and controlled implementation.
The first-order coefficient $C_{\rm pair}$ alone does not fix $K_p$ for
higher orders. Appendix~\ref{app:higher_order} illustrates this trade-off
for the six-qubit ${}^{6}$He Hamiltonian using exact matrix exponentiation
as a reference. This limited convergence test does not replace the
first-order estimates in Figs.~\ref{fig:Qresource_summary} and \ref{fig:QPE_fixed_accuracy}.

Randomized methods provide further possibilities. In qDRIFT~\cite{Campbell2019qDRIFT},
the sampled evolution count is governed by the coefficient one-norm
$\lambda'=\sum_j|h_j|$, rather than term count alone, so many weak
3N terms need not have the same cost implication as in a full Trotter sweep.
TE-PAI~\cite{Kiumi2025TEPAI} instead yields unbiased time-evolved
expectation-value estimators using random circuits, trading circuit depth
for sampling overhead. Their application to the eigenvalue algorithms
considered here requires an assessment of the relevant estimators and
repetition costs, which is beyond the present comparison.

Block-encoding-based approaches provide another route to long-time
evolution. In particular, constructions based on quantum singular value
transformation (QSVT)~\cite{Gilyen19} can replace the Trotterized
time-evolution operator in QKrylov and ODMD, and the qubitized walk
operator suggests polynomial Krylov or ODMD-like variants
(Appendix~\ref{app:qsvt_time_evolution}). We do not include these
variants in our estimates.

Symmetry tapering fixes the eigenvalues of independent commuting
Pauli $\mathbb{Z}_2$ symmetries, reducing qubit counts and potentially merging
Hamiltonian terms~\cite{Bravyi2017Tapering}.
Proton- and neutron-number parities provide candidates; further independent
symmetries must be checked for the chosen Hamiltonian and target sector.
Fixing full angular-momentum or isospin quantum numbers is not automatically
equivalent to such Pauli-symmetry tapering. Compact encodings of a restricted
configuration space, including Gray-code constructions, likewise require
separate estimates of operator and circuit costs~\cite{Steudtner2018Mappings}.
These possibilities are distinct from changing the untapered mapping
(Sec.~\ref{subsec:Encoding}).
The present estimates retain the full JW registers and do not quantify
these additional reductions.

\subsection{Toward sharper nuclear quantum benchmarks}

One immediate next step is to turn the present framework into a more systematic benchmark suite.
In practice, this means expanding the set of nuclear Hamiltonians, model spaces, and symmetry sectors,
while keeping the workflow sufficiently standardized that different algorithms and hardware assumptions can be compared on equal footing.
For each nucleus, a benchmark could specify the interaction, model-space
truncation, symmetry sector, and target observable. For a ground-state
or excitation energy, a reference calculation in the same space would
allow different algorithms to be compared at a common energy tolerance.
Repeating the calculation in successively larger spaces would then
test the convergence of the energy itself. The same distinction applies
to transition matrix elements, for which the transition operator and
the initial and final states must also be specified.

On the algorithmic side, Appendix~\ref{app:higher_order} tests
higher-order product formulas for short-time evolution in ${}^{6}$He.
Extending that test to full QPE circuits would show how the choice of
formula and its step count affect the T-gate comparison with qubitization
at fixed energy accuracy. For nucleus-specific ground-state estimates,
a complementary comparison would include the cost of preparing each
trial state and the number of repetitions needed at its achieved
ground-state overlap. This would put improvements in state preparation
and measurement grouping on a common footing: the total cost of an
energy estimate at a specified accuracy and success probability.

\subsection{Subspace and sampling methods}

For earlier hardware regimes, a promising direction is the class of
subspace methods that use quantum hardware to sample important
configurations, such as quantum selected configuration interaction
(QSCI)~\cite{QSCI} and sample-based quantum diagonalization
(SQD)~\cite{SQD_IBMRIKEN25}, followed by classical diagonalization of
the Hamiltonian in the sampled subspace. Their value must be assessed
against strong classical baselines, including selected-CI approaches
such as heat-bath CI and methods tailored to large-scale shell-model
calculations, such as the Monte Carlo shell model~\cite{MCSM_2001,
MCSM_2012, MCSM_2017} and the quasi-vacua shell
model~\cite{Shimizu_PRC21_QVSM}. Since subspace diagonalization is
available in
\texttt{NuclearToolkit.jl}~\cite{NuclearToolkit.jl,*Repo_NuclearToolkit.jl},
NuQuLib can provide an interface for testing QSCI- and SQD-type
workflows with quantum software such as Qiskit~\cite{Qiskit}.

\subsection{Broader extensions}

The framework developed here is not limited to the harmonic-oscillator shell-model setting.
Although this basis is the standard language of nuclear structure calculations,
there is growing interest in alternative formulations that may be advantageous for quantum simulation.
One example is lattice-based nuclear calculations,
including recent work based on pionless effective field theory on a lattice~\cite{Gu:2026} and
resource estimates for nuclear effective field theories on spatial
lattices~\cite{Watson2026,Spagnoli2026}.
Extending benchmark studies to such formulations would broaden the connection between quantum algorithms and the wider landscape of nuclear theory.

More broadly, an important long-term goal is to understand which combinations of physical formulation,
encoding, state preparation, and eigensolver are best matched to different nuclear problems.
The answer may differ between valence-space spectroscopy, no-core calculations, lattice formulations, and eventually reaction or scattering problems.
The purpose of the present work is therefore not to advocate a single algorithmic path,
but to establish a common starting point from which these alternatives can be compared in a controlled and reproducible way.

\section*{Acknowledgments}

This research was supported in part
by JSPS Grant-in-Aid for Scientific Research (Grant Nos.~JP22K14030, JP25H01511, 25K07294, 25K00995, 25K07330, 26H01394),
JST PRESTO (Grant No.~JPMJPR25F8),
JST ERATO (Grant No.~JPMJER2304),
RIKEN TRIP initiative (Nuclear Transmutation), and RIKEN Quantum.
ER is supported by the U.S. Department of Energy (DOE) under Contract No. DE-AC02-05CH11231, through the National Energy Research Scientific Computing Center (NERSC), an Office of Science User Facility located at Lawrence Berkeley National Laboratory. AB is supported by the U.S. Department of Energy (DOE) under Contract No. DE-AC05-00OR22725 through the Oak Ridge Leadership Computing Facility, an Office of Science User Facility at Oak Ridge National Laboratory.

This manuscript has been authored in part by UT-Battelle, LLC, under contract DE-AC05-00OR22725 with the US Department of Energy (DOE). The US government retains and the publisher, by accepting the article for publication, acknowledges that the US government retains a nonexclusive, paid-up, irrevocable, worldwide license to publish or reproduce the published form of this manuscript, or allow others to do so, for US government purposes. DOE will provide public access to these results of federally sponsored research in accordance with the DOE Public Access Plan (\url{http://energy.gov/downloads/doe-public-access-plan}).

We thank Takeshi Sato, Takumi Ogata, and Ritsuki Ito
for testing the early version of NuQuLib and giving us useful feedback.
We also thank Masaaki Kimura, Takashi Abe, and Noritaka Shimizu for useful discussions on
the results of NISQ simulations and possible future directions.

The authors declare no conflicts of interest.

\section*{Data Availability}

The NuQuLib code is available in the GitHub repository and the corresponding Zenodo archive~\cite{Repo_NuQuLib, *Zenodo_NuQuLib}, and the tutorial notebook generates Fig.~\ref{fig:QKrylov_ODMD_22O}. Additional benchmark data and plotting scripts are available in a separate GitHub repository~\cite{Repo_Data}.

\appendix

\section{T-gate count estimates for quantum algorithms}\label{app:Tformulas}

In this appendix, we summarize the formulas for T-gate count estimates
for the quantum algorithms discussed in this work.
The formulas below give the total T-gate count for one execution of each
distinct circuit. Table~\ref{tab:Qresources_Methods} includes the
number of repetitions per circuit, $N_{\mathrm{shot}}$.

\subsection{Quantum Phase Estimation}

In the repeated-power construction, the number of calls to the controlled
basic time-evolution operator is
\begin{align}
  \sum^{N_a-1}_{k=0} 2^k = 2^{N_a} - 1,
\end{align}
where $N_a$ is the number of ancilla qubits. 

For QPE, the required number of ancilla qubits can be estimated
from the target phase-estimation energy resolution $E_{\rm tol}$ and the time step $\delta t$ as
\begin{align}
N_a \approx \left\lceil \log_2 \left( \frac{2\pi}{|\delta t|\, E_{\rm tol}} \right) \right\rceil .
\end{align}
For the fixed-register estimates in Fig.~\ref{fig:Qresource_summary}(a,b),
we take $\delta t=10^{-2}\,\mathrm{MeV}^{-1}$, giving an energy interval
of width $2\pi/|\delta t|\simeq628$ MeV without phase aliasing.
The choices $E_{\rm tol}=1\,\mathrm{keV}$ and $1\,\mathrm{MeV}$
give $N_a=20$ and $10$, respectively.

The inverse quantum Fourier transform (QFT$^\dagger$) acts on the
phase register. We use the same conservative decomposition as in
Appendix~\ref{app:qubitized_QPE}: each controlled phase gate is counted
as three single-qubit rotations and Clifford gates. Thus
\begin{align}
 T_{\mathrm{QFT}^\dagger}\approx3\binom{N_a}{2}T_\epsilon.
\end{align}
This counts all rotations with the same synthesis model, without
optimizing special angles. For Fig.~\ref{fig:Qresource_summary}(a,b),
$T_\epsilon=100$; Fig.~\ref{fig:QPE_fixed_accuracy} sets the rotation precision from the
full-circuit synthesis budget. The QFT cost is included but is small
relative to repeated controlled evolution for the evaluated cases.
Approximate-QFT synthesis~\cite{Park:2025} could reduce it further.

The T-gate count for one QPE shot is
\begin{align}
T_{\mathrm{QPE}} & = T_{c\hat{U}} (2^{N_a} - 1) + T_{\mathrm{QFT}^\dagger},
\end{align}
where $ T_{c\hat{U}} $ is the T-gate count for the controlled version of the unit time-evolution gate,
which is naively twice the T-gate count for the uncontrolled version, i.e., $T_{\hat{U}}$.

\subsection{Quantum Krylov Subspace Method (QKrylov)}\label{sec:QKrylov}

For Fig.~\ref{fig:Qresource_summary}(a,b), we use the time-evolved Krylov vectors
\begin{align}
  \ket{\Phi_k}=e^{-i H k\delta t}\ket{\Phi_0},\qquad k=1,\dots,N_{\mathrm{iter}}.
\end{align}
For a pair $(m,n)$ with $m<n$, the two branches use $m$ and $n$
controlled evolution steps. Measuring both quadratures therefore costs
$2(m+n)$ controlled steps per setting. The finite-shot calculation in
Appendix~\ref{app:r21_fixed_sampling} includes the initial determinant
in the basis and applies the common evolution only once.

On the other hand, diagonal terms with $m=n$ can be measured by uncontrolled time evolution,
and thus the cost for each diagonal term is cheaper than the off-diagonal terms.
The number of calls for unit time evolution for diagonal terms
to measure $\bra {\Phi_m} H \ket{\Phi_m}$ is given as
\begin{align}
 N_{\mathrm{circ}} \sum^{N_{\mathrm{iter}}}_{m=1} m =  N_{\mathrm{circ}} \frac{N_{\mathrm{iter}}(N_{\mathrm{iter}}+1)}{2}.
\end{align}
Besides, measuring the overlap matrices $\tilde{N}$ can be realized as a special case of
the corresponding measurement for $\tilde{H}$. Hence, the cost for measuring $\tilde{N}$ is already included in the cost for measuring $\tilde{H}$.

Therefore the total number of controlled calls for all off-diagonal matrix elements is
\begin{align}
N_{\mathrm{circ}} \cdot 2 \sum_{m=1}^{N_{\mathrm{iter}}} \sum_{n=m+1}^{N_{\mathrm{iter}}} (m+n) = N_{\mathrm{circ}} (N^3_{\mathrm{iter}} - N_{\mathrm{iter}}).
  \label{eq:Ncall_Hnondiag}
\end{align}
For a fixed state and observable, independent repetitions reduce the
statistical standard error in proportion to the inverse square root of
the shot count. Propagation to the final eigenvalue error also depends
on the conditioning of the generalized eigenvalue problem.

As noted in Sec.~\ref{sec:quantum_algorithms}, our counts do not use the
Toeplitz structure of exact time evolution. The prescription in
Appendix~\ref{app:grouping_prescription} supplies $N_{\mathrm{circ}}$
for Fig.~\ref{fig:Qresource_summary}(a,b); the achievable reduction in
$N_{\mathrm{circ}}$ depends on the commutation structure of the
measured operators.

Summing the T-gate counts of all QKrylov measurement circuits, with one shot per circuit, gives
\begin{align}
T_{\mathrm{QKrylov}} 
& =
T_{\hat{U}}  N_{\mathrm{circ}} \frac{N_{\mathrm{iter}}(N_{\mathrm{iter}}+1)}{2} 
\nonumber \\
&\quad + T_{c\hat{U}} N_{\mathrm{circ}} (N^3_{\mathrm{iter}} - N_{\mathrm{iter}}),\nonumber \\
& = T_{\hat{U}} N_{\mathrm{circ}} \frac{ 4N^3_{\mathrm{iter}} + N^2_{\mathrm{iter}} - 3N_{\mathrm{iter}}}{2},
\end{align}
where we used the relation $T_{c\hat{U}} = 2 T_{\hat{U}}$ in the last line.

\subsection{Observable Dynamic Mode Decomposition (ODMD)}

ODMD requires only the autocorrelation samples $o_k=\langle\Phi_0|\Phi_k\rangle$. Together with the known value
$o_0=1$, these samples determine both Hankel matrices in Eq.~\eqref{eq:Hankel}.
The number of controlled time evolution gates required for ODMD is given as
\begin{align}
  2 \sum_{k=1}^{N_\mathrm{snap}} k = N_\mathrm{snap}(N_\mathrm{snap} + 1),
\end{align}
where $N_\mathrm{snap}$ is the number of snapshots in the ODMD method,
and the factor $2$ comes from measuring both real and imaginary parts of the overlap.
The T-gate count for ODMD is then given as
\begin{align}
T_{\mathrm{ODMD}} & = T_{c\hat{U}} \cdot N_{\mathrm{snap}}(N_{\mathrm{snap}} + 1).
\end{align}

The number of shots needed for a target accuracy is examined for two
nuclei in Sec.~\ref{sec:finite_shot} and
Appendix~\ref{app:r21_fixed_sampling}.

\section{Majorana-based evaluation of commutator coefficients and measurement settings}
\label{app:pair_bound}

\subsection{Pairwise commutator coefficient}
\label{app:pair_coefficient}

Our goal is to evaluate the pairwise commutator coefficient $C_{\rm pair}$
in the first-order Trotter error bound of Eq.~\eqref{eq:Bound_Pair}.
Each unordered pair of anticommuting Hamiltonian terms contributes the
product of their absolute coefficients. Testing all pairs directly
requires $O(L^2)$ commutation tests for $L$ terms. We compute the same
sum by accumulating weights over shared Majorana indices.
For each fermion mode $p$, define $\gamma_{2p}=a_p+a_p^\dagger$ and
$\gamma_{2p+1}=i(a_p^\dagger-a_p)$, so that
$\{\gamma_\mu,\gamma_\nu\}=2\delta_{\mu\nu}I$.
A particle-number-conserving $k$-body term contains $2k$ creation/annihilation
operators and therefore expands into even Majorana monomials of degree at most $2k$.
Each Jordan--Wigner Pauli term corresponds to a Hermitian even Majorana monomial
$G_A=i^{|A|/2}\prod_{\mu\in A}^{\rm ascending}\gamma_\mu$ of maximum degree $D=4$
for NN interactions, or $D=6$ with 3N interactions. This is a one-to-one relabeling
of the same Pauli decomposition, with unchanged absolute coefficients.
Even monomials anticommute if and only if $|A\cap B|$ is odd~\cite{Bravyi2010}.
For merged supports and weights $w_A=|h_A|$, define
\begin{align}
F_s(S)&=\sum_{A\supseteq S}w_A,\quad |S|=s,\\
M_s&=\frac12\left[\sum_{|S|=s}F_s(S)^2-\sum_A\binom{|A|}{s}w_A^2\right], \label{eq:Ms_Majorana}\\
C_{\rm pair}&=\sum_{s=1}^{D-1}(-2)^{s-1}M_s. \label{eq:Cpair_Majorana}
\end{align}
To see how the subset sums recover the contribution of each pair,
consider a fixed set $S$ of $s$ Majorana indices. Each term whose
support $A$ contains $S$ adds its weight $w_A$ to $F_s(S)$.
When $F_s(S)$ is squared, two distinct supports $A$ and $B$ containing
$S$ produce the cross term $2w_Aw_B$. The square also contains a
diagonal term $w_A^2$ for each support $A$ containing $S$.
Summing over all $S$ of size $s$ counts this diagonal term
$\binom{|A|}{s}$ times, which explains the subtraction in Eq.~\eqref{eq:Ms_Majorana}.
The factor of $1/2$ then counts each unordered pair once for each
subset $S$ that it shares.

A pair of supports sharing $k=|A\cap B|$ indices has exactly
$\binom{k}{s}$ shared subsets of size $s$. We may therefore write
\[
  M_s=\sum_{A<B}\binom{|A\cap B|}{s}\,w_Aw_B,
\]
where $A<B$ denotes an unordered pair of distinct supports.
This expression follows from the weighted intersection identity in
Eq.~(2.9) of Ref.~\cite{Bjorklund2008Intersection} by weighting each
support $B$ by $w_B$, summing over $B$, removing self-pairs, and
dividing by two.
For example, a pair sharing exactly two indices contributes
$2w_Aw_B$ to $M_1$ and $w_Aw_B$ to $M_2$. Its contribution to
$M_1-2M_2$ is therefore zero. A pair sharing just one index
contributes only to $M_1$ and retains its weight $w_Aw_B$.
For a general overlap size $k$, the binomial theorem gives
\[
  \sum_{s=1}^{k}(-2)^{s-1}\binom{k}{s}
  =\frac{1-(-1)^k}{2}.
\]
Thus Eq.~\eqref{eq:Cpair_Majorana} retains $w_Aw_B$ exactly once for each pair with odd
overlap, as required for the sum over anticommuting terms.
Equation~\eqref{eq:Cpair_Majorana} thus applies the inclusion--exclusion formula for
symmetric differences with weight $w_Aw_B$ assigned to each unordered pair.
Distinct supports of size at most $D$ cannot overlap
in $D$ indices, so the last possible contribution vanishes. Given the merged supports,
the unpartitioned cost of evaluating these moments is
$O(L2^D+\sum_{s=1}^{D-1}\binom{2N_q}{s})$ for $L$ terms, instead of $O(L^2)$ pair enumeration.

The moments can be accumulated as coefficient partitions are merged, avoiding
explicit pair enumeration. When the subset arrays do not fit in memory, their
index ranges are partitioned and the partial moment sums are combined.
This requires additional passes over the retained terms but leaves the identity unchanged.

For NN+3N, angular-momentum and isospin recoupling is factored into sparse
transforms of the canonical coupled matrices. Both parity blocks with the same
three-particle $(T_z,M)$ are merged together. A degree-six Majorana support fixes
this group: the sum of its six mode quantum numbers equals $(2T_z,2M)$.
Consequently, different groups have disjoint degree-six supports. Lower-degree
contractions are combined across all groups and with NN terms before the final cutoff.
All retained supports enter the moment sums; no sampling is used to evaluate
the commutator coefficient.

For the valence-space commutator benchmarks, we use CKpot ($p$)~\cite{CohenKurath}, USDB ($sd$)~\cite{USDB},
YSOX ($psd$)~\cite{Yuan2012}, GXPF1A ($pf$)~\cite{Honma2005},
SDPF-MU ($sd$--$pf$)~\cite{Utsuno2012}, and SDPFSDG-MU ($sd$--$pf$--$sdg$)~\cite{Utsuno2012PTP}.
The no-core benchmarks use EM500 NN inputs at $\hbar\omega=20$ MeV and an SRG resolution scale
$\lambda_{\rm SRG}=2.0\ \mathrm{fm}^{-1}$. The NN+3N sequence combines these NN
inputs with the supplied 3N interaction at the same SRG scale and uses
$E_{3\max}=3 e_{\max}$ throughout. The input files and coefficient conventions
used for the reported values are specified in the accompanying data.

\subsection{Measurement-setting bounds from Majorana pairings}
\label{app:grouping_prescription}

A perfect matching of the $2N_q$ Majorana labels defines one measurement
setting. We say that a matching covers a support $A$ when $A$ is a union
of its pairs; then $G_A=\pm\prod_{(a,b)\subseteq A}i\gamma_a\gamma_b$,
where the product runs over the pairs of the matching contained in $A$.
A signed permutation of the Majorana operators, implemented by a Clifford
circuit, maps the matching to the pairs $(2p,2p+1)$, for which
$i\gamma_{2p}\gamma_{2p+1}=-Z_p$. A single computational-basis measurement
then gives all pair eigenvalues $\pm1$ and hence the values of all
covered terms.

Table~\ref{tab:measurement_grouping} compares three greedy constructions.
Each starts from the same coefficients, ordered by decreasing absolute
value, after merging duplicate terms and applying a $10^{-8}$ MeV cutoff.
FC insertion tests commutation with a binary generator basis. Majorana
insertion tests compatibility with disjoint pairs; every resulting group
is also FC. We verify coverage and commutation for all assigned terms.
Different greedy choices can give more FC settings than a particular
Majorana construction, without contradicting the inclusion of the latter
in FC.

For a coefficient-independent cover, put $q=N_q$ and let $p$ be the
smallest prime satisfying $p\ge2q$ and $p\equiv3\pmod4$.
On $\mathbb F_p\cup\{\infty\}$, the map
\begin{align}
 f_{c,\delta}(x)=c+\frac{\delta}{x-c},
 \qquad c\in\mathbb F_p,
 \label{eq:majorana_field_pairing}
\end{align}
with $\delta$ a nonzero quadratic nonresidue is a fixed-point-free
involution, with $c$ paired to $\infty$.
It defines a perfect matching. Retain its physical--physical pairs on
the $2q$ Majorana labels and pair the remaining physical labels arbitrarily;
no auxiliary fermionic modes are required.
There are
\begin{align}
 B_2(q)=F(q)\equiv p(p-1)/2
 \label{eq:majorana_two_body_bound}
\end{align}
candidate settings. Every two-label support $\{u,v\}$ is contained in one
of them. The condition $f_{c,\delta}(u)=v$ reads $(u-c)(v-c)=\delta$.
The product $(u-c)(v-c)$ differs from the ratio $(v-c)/(u-c)$ by the
square factor $(u-c)^2$, and $c\mapsto(v-c)/(u-c)$ maps
$\mathbb F_p\setminus\{u\}$ onto $\mathbb F_p\setminus\{1\}$. Hence the
product is a nonresidue for $(p-1)/2$ values of $c$; each such $c$, with
$\delta=(u-c)(v-c)$, pairs $u$ with $v$.
For four distinct labels $u,v,w,t$, consider the three possible pairings.
For a pairing $(u,v),(w,t)$, the conditions $f_{c,\delta}(u)=v$ and
$f_{c,\delta}(w)=t$ read $(u-c)(v-c)=(w-c)(t-c)=\delta$. The quadratic
terms in $c$ cancel, and for $u+v\neq w+t$,
\[
 c=\frac{uv-wt}{u+v-w-t},\qquad
 \delta=\frac{N_{uv|wt}}{(u+v-w-t)^2},
\]
with the numerator $N_{uv|wt}=(u-w)(u-t)(v-w)(v-t)$. Thus $\delta$ is a
nonresidue exactly when $N_{uv|wt}$ is. Defining $N_{uw|vt}$ and
$N_{ut|vw}$ analogously for the other two pairings, one finds
\begin{gather*}
 N_{uv|wt}\,N_{uw|vt}\,N_{ut|vw}=-\Delta^2,\\
 \Delta=(u-v)(u-w)(u-t)(v-w)(v-t)(w-t),
\end{gather*}
which is minus a nonzero square.
Since $-1$ is a nonresidue, at least one numerator is a nonresidue.
A pairing with $u+v=w+t$ has $N_{uv|wt}=[(u-w)(v-w)]^2$, a square, and
cannot be that choice.
Thus one matching contains both pairs and covers the four-label support.
This gives an explicit $O(q^2)$ cover of all degree-2/4 Majorana operators,
consistent with the quadratic-setting approach of
Ref.~\cite{BonetMonroig2020Measurements}.

For degree six, divide the $2q$ labels into five balanced sets with sizes
$n_b$, where $\sum_b n_b=2q$ and $|n_b-n_{b'}|\le1$.
Every six-label support contains a pair in one of these sets.
Enumerate each such pair, fix it in the matching, and apply the
degree-2/4 construction to the remaining $2(q-1)$ labels.
Adding the original lower-degree cover gives
\begin{align}
 B_3(q)=F(q)+\left[\sum_{b=1}^{5}\binom{n_b}{2}\right]F(q-1).
 \label{eq:majorana_three_body_bound}
\end{align}
This is a constructive $O(q^4)$ setting upper bound, not an optimality
claim or a measured large-space group count. Small-space exhaustive
coverage checks provide an independent implementation test.
The same formulas are used for all QKrylov points in Fig.~\ref{fig:Qresource_summary}(a,b);
the table's greedy values are not substituted selectively into that figure.

\section{Block encoding and qubitized phase estimation}
\label{app:qubitization}

This appendix summarizes the block-encoding construction used for
qubitized QPE. We first introduce the linear combination of unitaries
(LCU)~\cite{Childs:2012} and the block encoding of the qubit
Hamiltonian, then the qubitized walk operator, and finally the cost
model used for Fig.~\ref{fig:QPE_fixed_accuracy}. Once the walk operator is available, phase
estimation can be applied to it directly and the Hamiltonian eigenvalue
recovered from the walk eigenphase, without first synthesizing
$e^{-iHt}$. QSVT-based time evolution, which is relevant for QKrylov and
ODMD, is summarized in Appendix~\ref{app:qsvt_time_evolution}.

\subsection{LCU block encoding}

We write the qubit Hamiltonian as a linear combination of Pauli unitaries,
\begin{equation}
    H = \sum_{\ell=1}^{N_P} a_\ell P_\ell ,
    \qquad
    a_\ell \geq 0 ,
\end{equation}
where $N_P$ is the number of Pauli terms. Here $P_\ell$ denotes a
Pauli word including its phase convention. In other words, any sign or
phase originally associated with a Pauli string is absorbed into
$P_\ell$, so that the LCU coefficients $a_\ell$ are non-negative.

We define the LCU normalization
\begin{equation}
    \lambda_H = \sum_{\ell=1}^{N_P} a_\ell.
    \label{eq:lambda_H}
\end{equation}
This normalization is a central quantity in all block-encoding-based
estimates below. It replaces the commutator-dependent product-formula
error analysis by an oracle-level cost model controlled by $\lambda_H$ and
by the costs of the data-loading and selection oracles.

The LCU block encoding is specified by the PREPARE and SELECT oracles,
\begin{align}
    \mathrm{PREP}|0\rangle
    &=
    \sum_{\ell=1}^{N_P}
    \sqrt{\frac{a_\ell}{\lambda_H}}
    |\ell\rangle \equiv \ket{\mathcal{L}}, \\
    \mathrm{SELECT}
    &=
    \sum_{\ell=1}^{N_P}
    |\ell\rangle \langle \ell|
    \otimes
    P_\ell.
\end{align}
The number of index-register qubits is
\begin{equation}
    N_A = \left\lceil \log_2 N_P \right\rceil.
\end{equation}
With
\begin{equation}
    U_{H/\lambda_H} =
    \mathrm{PREP}^{\dagger} \cdot \mathrm{SELECT} \cdot \mathrm{PREP},
\end{equation}
we obtain the block encoding
\begin{equation}
    (\langle 0| \otimes I)
    U_{H/\lambda_H}
    (|0\rangle \otimes I)
    =
    \frac{H}{\lambda_H}.
\end{equation}

The gate cost of this block encoding depends on the implementation of
PREPARE and SELECT. We therefore keep the corresponding T-gate counts as
\begin{equation}
    T_\mathrm{BE} =
    2T_{\rm PREP}
    +
    T_{\rm SELECT}.
\end{equation}
This notation is intentionally oracle-level. Different choices on how to 
implement oracles can lead to different constant factors and scaling behavior.

\subsection{Qubitized walk operator}
\label{app:Qubitization}
From the LCU construction above,  we define the prepared ancilla state as
\begin{equation}
    \ket{\mathcal{L}} = \mathrm{PREP}\ket{0},
\end{equation}
and the corresponding reflection as
\begin{equation}
    \mathcal{R}_{\mathcal{L}}
    =
    2\ket{\mathcal{L}}\bra{\mathcal{L}} \otimes I - I. \label{eq:reflection_L}
\end{equation}
The qubitized walk operator is then defined by
\begin{equation}
    W = \mathcal{R}_{\mathcal{L}}\,\mathrm{SELECT}. \label{eq:walk_operator}
\end{equation}

This form is equivalent to the more compact expression obtained in the computational basis of the ancilla register. Writing
\begin{equation}
    \Pi = \ket{0}\bra{0}\otimes I,
\end{equation}
and recalling that
\begin{equation}
    U_{H/\lambda_H}
    =
    \mathrm{PREP}^{\dagger}\,\mathrm{SELECT}\,\mathrm{PREP},
\end{equation}
one finds
\begin{equation}
    \mathrm{PREP}^{\dagger} W \,\mathrm{PREP}
    =
    (2\Pi-I)\,U_{H/\lambda_H}.
\end{equation}
Thus, the two forms are simply related by conjugation with PREP, and we use
$W = \mathcal{R}_{\mathcal{L}}\,\mathrm{SELECT}$ as the defining expression in what follows.

Let
\begin{equation}
    H\ket{\psi_k} = E_k\ket{\psi_k}.
\end{equation}
Then the walk operator has a two-dimensional invariant subspace
associated with $\ket{\psi_k}$, in which its eigenvalues are $e^{\pm i\theta_k}$,
with $\cos\theta_k = E_k /\lambda_H$.
Thus, the Hamiltonian eigenvalue is encoded in the eigenphase of the
walk operator:
\begin{equation}
    E_k = \lambda_H \cos\theta_k. \label{eq:walk_eigenphase}
\end{equation}

This observation gives the most direct block-encoding-based formulation of QPE,
and one
can apply QPE directly to the walk operator $W$.

\subsection{Qubitized QPE from the walk operator}
\label{app:qubitized_QPE}

For a comparison at a common absolute energy accuracy, both algorithms
encode the identity-shifted Hamiltonian
\begin{equation}
 H'=H-h_I I=\sum_{j=1}^{M}h_jP_j,\qquad
 \lambda'=\sum_{j=1}^{M}|h_j|.
\end{equation}
Here $M$ is the retained nonidentity Pauli-term count of the same benchmark
used to evaluate $C_{\rm pair}$, rather than the structural count in Table~\ref{tab:ModelSpace}.
The known identity coefficient is restored classically. Applying the walk
construction above to $H'/\lambda'$ gives
\begin{equation}
 E_k=h_I+\lambda'\cos\theta'_k.
\end{equation}
Thus subtracting the identity changes the walk normalization without
changing the desired absolute energy accuracy or $C_{\rm pair}$.
Accuracy here is relative to the retained, coefficient-screened benchmark
Hamiltonian; model-space and interaction uncertainties are not included.

We specify a target error $\Delta E$ for a spectral sample, and allocate
$\epsilon_{\rm PE}=\Delta E/2$ to phase estimation and
$\epsilon_H=\Delta E/2$ to the encoded-Hamiltonian approximation.
The latter is the product-formula spectral error for Trotterized QPE and
the coefficient-rounding error for qubitized QPE. These are additive
energy-error allocations, in MeV, rather than failure probabilities.
Costs are conditional on an eigenstate of the encoded operator; equal
usable target-state overlap is assumed when comparing the two methods.
Preparing a nuclear eigenstate and determining how either approximation
changes the ground-state overlap are outside this comparison.

For both methods we use conventional QPE with a uniform phase register
and an inverse QFT. Its two adjacent phase bins have combined probability
at least $8/\pi^2$, and either bin is within $2\pi/2^m$ of the encoded
phase. The sufficient phase-register sizes are therefore
\begin{align}
 m_T&=\left\lceil\log_2\frac{2\pi}{|\delta t|\epsilon_{\rm PE}}\right\rceil,
 &m_W&=\left\lceil\log_2\frac{2\pi\lambda'}{\epsilon_{\rm PE}}\right\rceil,\label{eq:qpe_common_bits}\\
 Q_T&=2^{m_T}-1,&Q_W&=2^{m_W}-1.
\end{align}
We take $\delta t=\pi/(2\lambda')$, in MeV$^{-1}$, so that
$\|H'\delta t\|\le\pi/2$ leaves a margin against phase aliasing.
The phase-to-energy conversion for the walk uses
$|\Delta E|\le\lambda'|\Delta\theta|$~\cite{Babbush:2018}.
In particular, a fixed phase-register size does not imply a fixed
absolute energy resolution across Hamiltonians with different $\lambda'$.

For Trotterized QPE we repeat the same base unitary
$V=S_1(\delta t/r)^r$ when constructing every power $V^{2^k}$.
The pairwise bound and the chord-to-phase relation give the sufficient choice
\begin{equation}
 r=\max\left(1,\left\lceil
 \frac{C_{\rm pair}\delta t^2}
 {2\sin(|\delta t|\epsilon_H/2)}\right\rceil\right).
 \label{eq:qpe_common_steps}
\end{equation}
Indeed, a unitary-norm error $b\le2$ bounds the distance to an exact
eigenphase by $2\arcsin(b/2)$; the consistent phase branch then bounds the
corresponding spectral energy error by $2\arcsin(b/2)/|\delta t|$.
This is a statement about spectral accuracy, not preservation of individual
eigenvectors. In the small-error limit it reduces to
$r\sim C_{\rm pair}|\delta t|/\epsilon_H$.
The pairwise bound provides sufficient steps and can overestimate the
steps needed in practice, because it discards commutator cancellations.

For the walk oracle we use coherent alias sampling and unary-iteration
SELECT~\cite{Babbush:2018,Uvarov:2026}.
We pad the $M$ entries with zero weights to $L=2^\ell$, where
$\ell=\lceil\log_2 M\rceil$. This explicit construction prepares the
uniform alias register with exact Hadamard gates; it requires no
synthesized arbitrary rotations inside PREPARE.
The unused SELECT entries act as the identity.
Normalized coefficient probabilities are rounded while preserving their
sum, with individual error at most $1/(L2^\mu)$.
Consequently the encoded Hermitian Hamiltonian $\widetilde{H}'$ satisfies
\begin{equation}
 \|\widetilde{H}'-H'\|\le\lambda'2^{-\mu}\le\epsilon_H,
 \qquad
 \mu=\left\lceil\log_2\frac{\lambda'}{\epsilon_H}\right\rceil.
 \label{eq:qpe_coefficient_precision}
\end{equation}
The signed Pauli operators are selected exactly; finite coefficient
precision is included in the energy budget, not accumulated as an
independent error on every oracle call.

With the QROM choice $q=1$ and the four-T, measurement-assisted Toffoli
convention of Ref.~\cite{Uvarov:2026}, the oracle model is
\begin{align}
 T_{\rm PREP}&=4L+4\mu+12\ell,\nonumber\\
 T_{\rm SELECT}&=4L-4,\qquad T_\Pi=4(\ell-1),\nonumber\\
 T_{cW}&=2T_{\rm PREP}+T_{\rm SELECT}+T_\Pi\nonumber\\
       &=12L+8\mu+28\ell-8.
 \label{eq:qpe_walk_cost}
\end{align}
This is the cost of a controlled walk call, so no additional factor of
two is applied. Controlled Pauli payloads are Clifford operations;
their weights affect Clifford cost but not this SELECT T-gate count, including
for the NN+3N Hamiltonians. The oracle ancilla count in this model is
$a_W=2\mu+3\ell+1$, separate from the phase register $m_W$ and the
$N_q$ system qubits. These are logical resource estimates; Clifford
runtime, physical error correction, and compilation optimizations are
not included. Power-of-two padding is counted in the oracle cost.

We also impose a common synthesis-error budget
$\epsilon_{\rm circ}=10^{-3}$ in operator norm for each complete QPE
circuit. A conservative inverse-QFT decomposition uses
$g(m)=3m(m-1)/2$ single-qubit rotations. The numbers of synthesized
rotations and their cost estimates are
\begin{align}
 n_R^T&=2MrQ_T+g(m_T),& n_R^W&=g(m_W),\nonumber\\
 \epsilon_R^A&=\epsilon_{\rm circ}/n_R^A,&
 T_R^A&=\left\lceil3\log_2(1/\epsilon_R^A)\right\rceil.
\end{align}
Here $A\in\{T,W\}$ labels the two algorithms.
The factor two in $n_R^T$ implements each controlled Pauli rotation
with two synthesized rotations. Telescoping the gate errors bounds
the circuit-norm error by $\epsilon_{\rm circ}$. This reduces the
conditional success probability by at most $2\epsilon_{\rm circ}$,
leaving it above $0.80$ for either method. Synthesis error is thus
accounted for as a probability penalty separately from the energy budget.
The logarithmic synthesis formula is a resource model, not an explicit
compilation of these circuits.
For $N_{\rm shot}$ preparations, the resulting estimates are
\begin{align}
 T_{\rm QPE}^{T}&=N_{\rm shot}\,n_R^TT_R^T,\nonumber\\
 T_{\rm QPE}^{W}&=N_{\rm shot}
   [Q_WT_{cW}+T_{\rm PREP}+g(m_W)T_R^W].
 \label{eq:qpe_common_cost}
\end{align}
The extra PREPARE initializes the walk ancillas; it does not prepare the
nuclear eigenstate. Figure~\ref{fig:QPE_fixed_accuracy} uses $N_{\rm shot}=1$ and includes
inverse-QFT costs. Its panels (a,b) show absolute T-gate counts at
$\Delta E=100$ keV, while panel (c) shows ratios at three target
accuracies. Ignoring integer rounding and these small additive
costs, the ratio scales as
\begin{equation}
 \frac{T_{\rm QPE}^{T}}{T_{\rm QPE}^{W}}
 \sim\frac{C_{\rm pair}}{\lambda'\epsilon_H}
       \frac{2M T_R^T}{T_{cW}}.
\end{equation}
All precision allocations and oracle costs in the plotted results are
specified explicitly; no freely varied prefactor is used.

\section{QSVT-based time evolution}
\label{app:qsvt_time_evolution}

\begin{table*}
\centering
\caption{Hamiltonian coefficient one-norms, including identity, and estimated even polynomial degrees
at $t=0.25\,\mathrm{MeV}^{-1}$ for the indicated simulation errors.
Input and degree conventions are specified in Appendix~\ref{app:qsvt_time_evolution}.
\label{tab:Trotter_Error_QSVT}}
\begin{ruledtabular}
\begin{tabular}{lrrrr}
 & & & \multicolumn{2}{c}{$d$ for $\epsilon_{\rm sim}$} \\
\cline{4-5}
Model space / interaction & $N_q$ & $\lambda_H$ (MeV) & $10^{-7}$ & $10^{-4}$ \\
\hline
$p$ (CK) & 12 & $2.62\times10^{2}$ & 106 & 100 \\
$sd$ (USDB) & 24 & $1.06\times10^{3}$ & 376 & 370 \\
$psd$ (YSOX) & 36 & $2.69\times10^{3}$ & 932 & 926 \\
$pf$ (GXPF1A) & 40 & $2.32\times10^{3}$ & 804 & 798 \\
$sd$--$pf$ (SDPF-MU) & 64 & $8.91\times10^{3}$ & 3,044 & 3,038 \\
$sd$--$pf$--$sdg$ (SDPFSDG-MU) & 124 & $6.35\times10^{4}$ & 21,594 & 21,586 \\
\hline
NN, $e_{\max}=0$ & 4 & $3.49\times10^{1}$ & 24 & 20 \\
NN, $e_{\max}=1$ & 16 & $6.83\times10^{2}$ & 250 & 242 \\
NN, $e_{\max}=2$ & 40 & $6.67\times10^{3}$ & 2,284 & 2,278 \\
NN, $e_{\max}=3$ & 80 & $3.96\times10^{4}$ & 13,476 & 13,470 \\
NN, $e_{\max}=4$ & 140 & $1.70\times10^{5}$ & 57,732 & 57,724 \\
NN, $e_{\max}=5$ & 224 & $5.79\times10^{5}$ & 196,836 & 196,828 \\
NN, $e_{\max}=6$ & 336 & $1.67\times10^{6}$ & 566,746 & 566,738 \\
NN, $e_{\max}=7$ & 480 & $4.23\times10^{6}$ & 1,436,354 & 1,436,348 \\
NN, $e_{\max}=8$ & 660 & $9.71\times10^{6}$ & 3,298,072 & 3,298,064 \\
NN, $e_{\max}=9$ & 880 & $2.06\times10^{7}$ & 7,001,448 & 7,001,442 \\
\hline
NN+3N, $e_{\max}=1$ & 16 & $7.98\times10^{2}$ & 288 & 282 \\
NN+3N, $e_{\max}=2$ & 40 & $1.56\times10^{4}$ & 5,326 & 5,320 \\
NN+3N, $e_{\max}=3$ & 80 & $2.32\times10^{5}$ & 78,782 & 78,776 \\
NN+3N, $e_{\max}=4$ & 140 & $2.23\times10^{6}$ & 759,346 & 759,340 \\
\end{tabular}
\end{ruledtabular}
\end{table*}

QSVT~\cite{Gilyen19,Martyn:2021} provides a block-encoding-based method to
synthesize the time-evolution operator itself. This is relevant for
algorithms such as QKrylov and ODMD, where the basic objects are
time-evolved states or time-dependent observables, and can be used also for QPE.

For $t>0$, write $X=H/\lambda_H$ and $\tau=\lambda_H t$. The polynomial
approximation used for Hamiltonian simulation satisfies
\begin{equation}
    P_d(X)\approx\exp(-i\tau X)=\exp(-iHt).
\end{equation}
In the standard QSVT construction, the even cosine and odd sine
polynomials are implemented by separate phase sequences and combined
coherently~\cite{Martyn:2021}. For an even degree $d$,
a representative sequence for one parity component is
\begin{equation}
    U_d
    =
    \left[
    \prod_{k=1}^{d/2}
    \Pi_{\phi_{2k-1}}
    U_{H/\lambda_H}^\dagger
    {\Pi}_{\phi_{2k}}
    U_{H/\lambda_H}
    \right]
    \Pi_{\phi_{d+1}},
\end{equation}
where 
$\{\phi_j\}_{j=1}^{d+1}$ are phase angles determined numerically~\cite{Chao:2020,Dong:2021}. The phase factors
are implemented through projector-controlled phase rotations of the
form
\begin{equation}
    \Pi_\phi
    =
    \exp\left[
    i\phi(2\Pi-I)
    \right],
\end{equation}
with $\Pi=|0\rangle\langle 0|\otimes I$ denoting the projector onto
the signal subspace. In a circuit implementation, such a projector
phase can be decomposed into multi-controlled operations and a
single-qubit $Z$ rotation, for example schematically as
\begin{equation}
    \Pi_\phi
    =
    C_{\Pi}X\, e^{-i\phi Z}\, C_{\Pi}X ,
\end{equation}
up to convention-dependent global phases. Thus, a degree-$d$ QSVT
approximation requires $d$ applications of the block-encoding unitary
or its inverse, and $d+1$ phase rotations.

The degree $d$ must be chosen based on the normalized simulation time $\lambda_H t$
and the target approximation error $\epsilon_{\rm sim}$. For the
Jacobi--Anger-type polynomial approximations commonly used for
Hamiltonian simulation~\cite{Berry2015}, one may express the truncation error in a form
such as~\cite{Low:2019},
\begin{equation}
    \epsilon_{\rm sim}
    \lesssim
    \frac{4(\tau/2)^d}{d!}
    \le
    \sqrt{\frac{8}{\pi}}
    \left(\frac{e\tau}{2d}\right)^d ,
\end{equation}
where the last bound uses $d!\ge\sqrt{2\pi d}(d/e)^d$
and $d\ge 1$. Equating this upper estimate to $\epsilon_{\rm sim}$
gives the continuous degree estimate
\begin{equation}
\begin{aligned}
    d_{\rm est} &= \frac{\tau}{2}\,
       \exp\!\left[1+W\!\left(\frac{b}{e\tau}\right)\right],
       \\
    b &= \log(8/\pi)-2\log\epsilon_{\rm sim}.
\end{aligned}
\end{equation}
Here $W$ is the principal Lambert function. The estimate scales as
$O(\tau+\log(1/\epsilon_{\rm sim}))$.

Table~\ref{tab:Trotter_Error_QSVT} lists the coefficient one-norms for
six valence spaces, NN spaces through $e_\mathrm{\max}=9$, and NN+3N spaces
through $e_\mathrm{\max}=4$. The input conventions are those of
Table~\ref{tab:measurement_grouping} and Appendix~\ref{app:pair_bound};
the six valence interactions are the same as in Fig.~\ref{fig:Trotter_error},
including SDPF-MU~\cite{Utsuno2012} and SDPFSDG-MU~\cite{Utsuno2012PTP}. The NN and NN+3N entries at a given
$e_\mathrm{\max}$ act on the same number of system qubits, with $E_{3\max}=3e_\mathrm{\max}$ for NN+3N.
Here the known identity coefficient is retained:
$\lambda_H=|h_I|+\lambda'$, whereas the QPE comparison in
Appendix~\ref{app:qubitized_QPE} uses $\lambda'$ after removing it.
These are coefficient sums over the full qubit Hamiltonian, not nuclear
binding energies or norms restricted to a fixed particle-number sector.

For $t=0.25\,\mathrm{MeV}^{-1}$, the listed degrees are obtained by
rounding the Lambert-function estimate upward to the next even integer
and checking the factorial expression above. The degrees use the
unrounded coefficient sums; $\lambda_H$ is displayed to three significant
digits. These estimates illustrate the dependence on normalization and
simulation error, without constructing optimized QSVT phase sequences.

The T-gate count for one degree-$d$ QSVT sequence can therefore be written as
\begin{equation}
    T_{\rm QSVT}(t,\epsilon_{\rm sim})
    \simeq
    d(\lambda_H t,\epsilon_{\rm sim}) T_\mathrm{BE}
    +
    (d+1)T_{\Pi_\phi},
\end{equation}
where $T_\mathrm{BE}$ is the cost of one block-encoding query and $T_{\Pi_\phi}$
is the cost of one projector-controlled phase rotation. A full simulation
also accounts for combining the parity components and any amplitude amplification. In terms of
the LCU oracles,
\begin{equation}
    T_\mathrm{BE}
    \simeq
    2T_{\rm PREP}
    +
    T_{\rm SELECT}.
\end{equation}
The projector-controlled phase cost may be expressed schematically as
\begin{equation}
    T_{\Pi_\phi}
    =
    2 T_{\rm proj}
    +
    T_R(\epsilon_R),
\end{equation}
where $T_{\rm proj}$ is the cost of implementing the control on the
signal subspace and $T_R(\epsilon_R)$ is the T-gate count required to
synthesize the single-qubit rotation $e^{-i\phi Z}$ to precision
$\epsilon_R$. In the numerical estimates in the main text, a
representative value $T_R\simeq 100$ is used for high-precision
single-qubit rotations.

If the QSVT time-evolution operator itself must be controlled, as in
Hadamard-test-type measurements of overlaps or off-diagonal matrix
elements, one should use a controlled cost
\begin{equation}
    T_{c{\rm QSVT}}(t,\epsilon_{\rm sim})
\end{equation}
rather than assuming that the control overhead is simply additive. In
general, controlling the QSVT sequence may change the cost of the
block-encoding queries, the projector-controlled phase rotations, and
the rotation synthesis. Therefore, a safe oracle-level expression is
\begin{equation}
    T_{c{\rm QSVT}}(t,\epsilon_{\rm sim})
    \simeq
    d(\lambda_H t,\epsilon_{\rm sim}) T_{cU}
    +
    (d+1)T_{c\Pi_\phi},
\end{equation}
where $T_{cU}$ and $T_{c\Pi_\phi}$ denote the implementation-dependent
controlled costs. Depending on the chosen circuit construction, the
controlled version may be close to an additive overhead, or it may
effectively multiply part of the cost, for example when data loading,
multi-controlled operations, or controlled rotations must be promoted
to their controlled versions.

For QKrylov, QSVT-based Hamiltonian simulation can replace the product-formula implementation of the Krylov states.
This replacement changes the coherent cost of preparing the
time-evolved states, but it does not remove the measurement overhead
associated with estimating the reduced Hamiltonian and overlap matrices,
Eqs.~\eqref{eq:Hmatrix} and~\eqref{eq:Nmatrix}.
Thus, at the oracle level, the product-formula time-evolution cost in
the main-text QKrylov estimate can be replaced by
$T_{\rm QSVT}(t,\epsilon_{\rm sim})$, or by
$T_{c{\rm QSVT}}(t,\epsilon_{\rm sim})$ when controlled time evolution
is required.

For ODMD, the real and imaginary parts of the autocorrelation at each
nonzero time $t_k$ are measured in separate circuits. Each uses one
controlled QSVT evolution. With $N_{\rm shot}$ repetitions per circuit,
the T-gate count for these evolutions is estimated as
\begin{equation}
    T_{\rm ODMD}^{\rm QSVT}
    \simeq
    2N_{\rm shot}
    \sum_{k=1}^{N_{\rm snap}}
    T_{c{\rm QSVT}}(t_k,\epsilon_{\rm sim}).
\end{equation}

\subsection{Possible extensions: block-encoding-based Krylov and ODMD variants}
\label{app:block_encoding_extensions}

The block-encoding framework also suggests possible extensions beyond
the direct walk-operator QPE and QSVT-based time evolution discussed
above. In particular, one may use the block encoding, or the associated
qubitized walk operator, to construct Krylov-type or ODMD-like
sequences without relying directly on product-formula time evolution.

A related direction has already been explored in the context of quantum
Lanczos methods based on block encodings, where Krylov subspaces are
constructed using polynomial functions of the Hamiltonian rather than
real-time evolution alone~\cite{Kirby:2023}. In the present setting,
the qubitized walk operator provides a natural polynomial structure.
Since its eigenphases satisfy Eq.~\eqref{eq:walk_eigenphase},
powers of the walk operator generate phase factors $e^{\pm in\theta_k}$. Through the identity
\begin{equation}
    \cos(n\theta)=T_n(\cos\theta),
\end{equation}
where $T_n$ is the Chebyshev polynomial, such sequences can be
interpreted as probing polynomial information in $H/\lambda_H$. This
suggests possible polynomial Krylov bases or ODMD-like time series based
on $W^n$, from which the eigenvalues could be recovered through Eq.~\eqref{eq:walk_eigenphase}.
A detailed analysis of the numerical stability, measurement cost, and
state-preparation requirements of these variants is beyond the scope of
this work.

In summary, block encoding offers several alternatives to
product-formula time evolution: walk-operator QPE
(Appendix~\ref{app:qubitization}), QSVT-based Hamiltonian simulation,
and polynomial Krylov or ODMD-like constructions. Resource estimates
for block-encoding-based QKrylov and ODMD remain open.

\section{Finite-shot ground-state energy estimates for $^{6}$He and $^{22}$O}
\label{app:r21_fixed_sampling}

Here we provide the details of the finite-shot ground-state energy estimates for $^{6}$He and $^{22}$O
summarized in Sec.~\ref{sec:finite_shot}.
We used the neutron $p$ shell with the CKpot interaction for
$^{6}$He~\cite{CohenKurath} and the neutron $sd$ shell with USDB for $^{22}$O~\cite{USDB}.
The initial determinant occupied the $0p_{3/2}$ orbitals with
$m_j=\pm 1/2$ in $^{6}$He and the full $0d_{5/2}$ subshell in $^{22}$O.
The squared ground-state overlaps were 0.38 and 0.75, respectively.
Exact diagonalization in the corresponding $M_J=0$ valence sectors
gave $E_0=-3.910$ and $-34.498$~MeV relative to the inert cores.
For time evolution we used the same ordered Jordan--Wigner expansion across methods
within each nucleus (105 and 1,690 nonidentity terms, respectively),
with four first-order Trotter substeps per interval $\delta t$.
We restored the identity shift classically and sampled outcomes from the ideal circuits.

We scored all three methods against the window $[E_0-\delta,E_0+\delta]$,
with $\delta=100$~keV. The gaps to the next level, 4.54 and 3.16~MeV for
$^{6}$He and $^{22}$O, place the excited states outside this window in each sector.
For QPE we chose the phase-register sizes $N_a$ to give energy bins narrower
than $\delta$ (49.1 and 41.4~keV, respectively). Even at this resolution,
the readout samples excited states, and the finite register gives some
readouts below the ground-state window. A single such readout can set the
sample minimum. We therefore looked for a low percentile that would
select the ground-state peak in both nuclei.

To choose that percentile we used the known $E_0$ during calibration.
For one decoded energy $X$, let $p_-=\Pr(X<E_0-\delta)$ and
$p_+=\Pr(X\le E_0+\delta)$. The fraction of readouts below the window
was $0.014\%$ for $^{6}$He and $2.37\%$ for $^{22}$O. Including
readouts in the window brought these fractions to $38.22\%$ and
$72.15\%$, respectively. Thus the median lies above the window for $^{6}$He,
while several lower percentiles lie in the window for both nuclei. We chose
one of these, the 25th percentile, as a common readout.
To return one estimate we sorted $n$ new energies and took the
$k$th smallest, with $k=\lceil n/4\rceil$. This step uses only the readouts.

For this estimate to succeed, at least $k$ shots must fall at or below
the upper edge, and fewer than $k$ may fall below the lower edge.
Independence of the shots gives the predicted success probability
\[
 \Pr\!\left[\mathrm{Bin}(n,p_+)\ge k\right]
 -\Pr\!\left[\mathrm{Bin}(n,p_-)\ge k\right],
\]
where $\mathrm{Bin}(n,p)$ denotes a binomial count. We chose a $98.5\%$
calibration threshold to leave a margin above the common $95\%$ validation
target. The first budgets to meet that threshold were 56 shots for $^{6}$He
(the 14th readout in sorted order, $98.72\%$ predicted success) and seven
for $^{22}$O (the second readout, $98.66\%$). Six shots already gave
$98.44\%$ predicted success in $^{22}$O. The selected budgets thus depend
on both the readout rule and the calibration threshold.

For ODMD and QKrylov we chose the number of shots $n$ per circuit as a
power of two for each method and nucleus. Since both methods combine
measurements from many circuits, we evaluated each allocation using the
error in the reconstructed ground-state energy.
ODMD sampled both quadratures of the autocorrelation at $K$ nonzero times
and reused each measurement wherever its time appeared in the Hankel matrices.
This gave $2K$ distinct circuits. For example, in
$^{22}$O we sampled each quadrature at 64 times with 16 shots per circuit,
giving $2\times64\times16=2{,}048$ executions per estimate.
For QKrylov we grouped the retained terms of each Hamiltonian by greedy Majorana pairing,
obtaining $G=16$ groups for $^{6}$He and $G=114$ for $^{22}$O.
Each setting measures products of commuting bilinears $i\gamma_a\gamma_b$ from a single pairing.
A Clifford basis change maps these observables to signed products of $Z$ operators.
For a basis of $m$ Krylov states including the unevolved determinant,
each group requires $m$ diagonal measurements and two quadratures for each distinct off-diagonal pair,
for a total of $Gm^2$ circuits.
Sampling the joint ancilla and group outcomes supplied both the Hamiltonian elements and,
from the ancilla marginals, the overlaps. 
This retains the covariance of observables measured in the same shot.

With $n$ shots per setting, we retained ODMD singular values above $\sqrt{2L/n}$
and QKrylov overlap eigenvalues above $\sqrt{2m(m-1)/(Gn)}$,
where $L$ counts nonzero-time entries, including repeats,
in the unshifted Hankel matrix of $K/2$ rows.
The energy estimate was the lowest energy inferred from the retained ODMD phases
or the lowest generalized eigenvalue in the retained overlap subspace.

To test whether the selected budgets met the common success target,
we simulated 1,200 independent energy estimates for each row of
Table~\ref{tab:r21_fixed_sampling}. Each estimate contributed one success
if it fell inside the target window. The lowest observed success count was
1,181 for ODMD in $^{22}$O. Even for this protocol the one-sided lower
confidence bound on the success probability was $97.33\%$, above the
$95\%$ target (exact binomial bounds with simultaneous $95\%$ coverage
across the six protocols using a Bonferroni correction).
At the tested allocations, the 95th percentiles of the absolute energy error
in $^{6}$He and $^{22}$O were 66.3 and 81.5~keV for ODMD, respectively,
and 58.7 and 60.1~keV for QKrylov.

The longest circuits contain $r(2^a-1)$, $rK$, or $r(m-1)$ sequential substeps for QPE,
ODMD, or QKrylov, respectively.
To compare time-evolution work over all shots, we use $W=N_u+2N_c$,
counting uncontrolled and controlled Pauli rotations with weights one and two.
For Krylov indices $i\leq j$, the first $i$ intervals are common to both branches
and the remaining $j-i$ are controlled. Each ratio uses the QPE value for the same nucleus.
This metric excludes state preparation, rotation synthesis, the inverse QFT,
and the Clifford readout circuits. The reported budgets are examples that
met the accuracy target with the chosen determinants, estimators, and
uniform shot allocations. The minimum budgets for these systems remain open.

\section{Quantum projection filter}
\label{app:qpf}

This appendix summarizes the QPF of Ref.~\cite{Stetcu:2023} for
reference. It is not used in the resource estimates or demonstrations
of this work.

From a circuit perspective, the filtering operation can be understood within the general block-encoding framework. Given a trial state $\ket{\psi_T}$ with nonzero overlap with the ground state of a Hermitian Hamiltonian $H$, a block encoding of a function $f_i(H)$ implements
\begin{equation}
\ket{0}_a\ket{\psi_T}\rightarrow \ket{0}_af_i(H)\ket{\psi_T}+\cdots    
\end{equation}
where the subscript \(a\) denotes the ancilla register. 
Post-selecting the ancilla measurement outcome $0$ yields
\begin{equation}
\frac{f_i(H)\ket{\psi_T}}{\sqrt{p_i}}    
\end{equation}
with the success probability
$p_i=\bra{\psi_T}f^2_i(H)\ket{\psi_T}$. Applying a sequence $K$ of such operations we arrive at a final state
\begin{equation}
    \ket{\phi}=\frac{\prod_{i=1}^Kf_i(H)\ket{\psi_T}}{\sqrt{p_{tot}}}
\end{equation}
where $p_{tot}=\bra{\psi_T}\prod_{i=1}^K f_i(H)^2\ket{\psi_T}$.
Using the single-ancilla block encoding derived from $e^{iY\otimes (tH+\delta\mathbf{1})}$ we obtain a filter function of the form $f_i(H)=\cos(t_iH+\delta_i)$. Recalling that the trial state can be expanded as a sum of the Hamiltonian eigenstates as
 \begin{equation}
     \ket{\psi_T}=\sum_n c_n\ket{\Phi_n}
 \end{equation}
 where \(H\ket{\Phi_n}=E_n\ket{\Phi_n}\), the ground-state overlap is
\begin{equation}
    \alpha = |\braket{\Phi_0|\psi_T}| = |c_0|.
\end{equation}
Here \(\ket{\Phi_0}\) denotes the true ground state of the Hamiltonian, while the state \(\ket{0}_a\) used above denotes the ancilla qubit state used for post-selection.
We obtain that the overlap between the filtered state and the ground state can be written as
 \begin{equation}
     \label{eq:qpf_overlap}
     \lvert\bra{\Phi_0}\phi\rangle\rvert^2=\frac{\lvert c_0\rvert^2\prod_i\cos^2(t_iE_0+\delta_i)}{\sum_n \lvert c_n\rvert^2\prod_i\cos^2(t_iE_n+\delta_i)}\, ,
 \end{equation}
 where $E_n$ are the eigenenergies of $H$.

Within this construction the cost of implementing each filtering step is determined by the implementation of the Hamiltonian evolution entering
$e^{iY\otimes(t_iH+\delta_i\mathbf{1})}$.
Writing the Hamiltonian as
\begin{equation}
    \hat{H}=\sum_{j=1}^{N_{\hat{H}}} a_j P_j, \ \lambda_H \equiv \sum_{j=1}^{N_{\hat{H}}}|a_j|
\end{equation}
where \(P_j\) are Pauli strings, the implementation cost depends on the selected Hamiltonian-simulation method and, in general, on the total filtering time, the target simulation accuracy, and Hamiltonian-dependent quantities such as $N_{\hat H}$, 
and the coefficient one-norm $\lambda_H$ defined in Eq.~\eqref{eq:lambda_H}.
The first-order Trotter step approximation error is due to the noncommutativity of the different Pauli
strings, as analyzed in Sec.~\ref{sec:pairwise_bounds}, and decreases as the time step \(\delta t\) is reduced, while resources for block encoding based methods will depend on $\lambda_H$. 

In controlled filtering operations, any known identity contribution must be included in the ancilla phase or absorbed into the adjustable filter phase $\delta_i$.

\section{Higher-order product-formula benchmark}
\label{app:higher_order}

We compare first-, second-, and fourth-order product formulas for the
six-qubit neutron-space CKpot Hamiltonian used in the ${}^{6}$He QPE
demonstration. After removing the known identity contribution, the
Hamiltonian $H'=H-h_II$ contains $M=105$ nonidentity Pauli terms.
We use the same evolution interval $t=0.2\,\mathrm{MeV}^{-1}$ as that
demonstration and compute $U(t)=e^{-itH'}$ by dense matrix exponentiation.
For each order $p$, the approximate evolution is
$V_p(t,r)=S_p(t/r)^r$ with $r=1,2,4,8,16,32$.
The error is $\|V_p(t,r)-U(t)\|_2$, the operator norm on the full
$64$-dimensional qubit space. Individual Pauli rotations are not projected
into the two-neutron sector, since they need not preserve it separately.

The product formulas retain NuQuLib's Pauli-term order and use Qiskit's
Suzuki construction with coefficients larger than $10^{-8}$ MeV in magnitude.
The matrix of the explicitly synthesized circuit is evaluated; the
abstract evolution gate itself represents the exact exponential.
An independent recursive product of Pauli exponentials checks the circuit
implementation. No rotation-synthesis error or measurement noise is added.

\begin{figure*}
\centering
\includegraphics[width=0.97\textwidth]{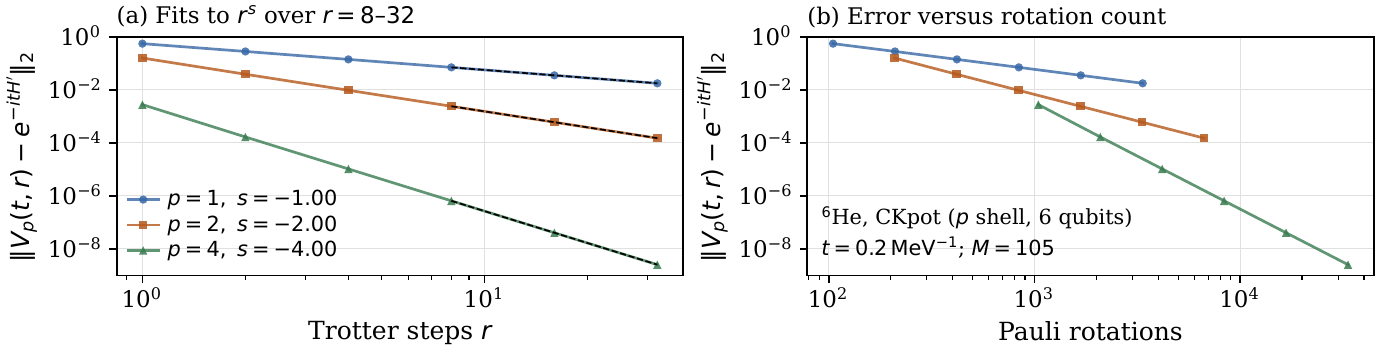}
\caption{Product-formula errors for the six-qubit ${}^{6}$He CKpot
Hamiltonian at $t=0.2\,\mathrm{MeV}^{-1}$, relative to exact matrix
exponentiation. The full-space operator-norm error is plotted against
(a) Trotter steps and (b) Pauli rotations, for orders $1$, $2$, and $4$.
Dashed segments in (a) show fits over $r=8$--$32$.
\label{fig:higher_order}}
\end{figure*}

Figure~\ref{fig:higher_order} shows convergence with both step count and
Pauli-rotation count. Fits over $r=8,16,32$ give powers of approximately
$-1.00$, $-2.00$, and $-4.00$, consistent with the expected orders.
The respective per-step counts are $105$, $209$, and $1045$ rotations,
using the convention specified in Sec.~\ref{subsec:extensions}.
The cost comparison shows why the extra work per step must be weighed
against the accuracy gain. These are short-time unitary errors, rather
than QPE energy errors or complete T-gate budgets.
An exhaustive study over nuclear model spaces, interactions, term
orderings, and target precisions is beyond the scope of this work.

\section{Supplementary VQE workflow example}
\label{app:vqe_demo}

\begin{figure*}
\centering
\includegraphics[width=0.99\textwidth]{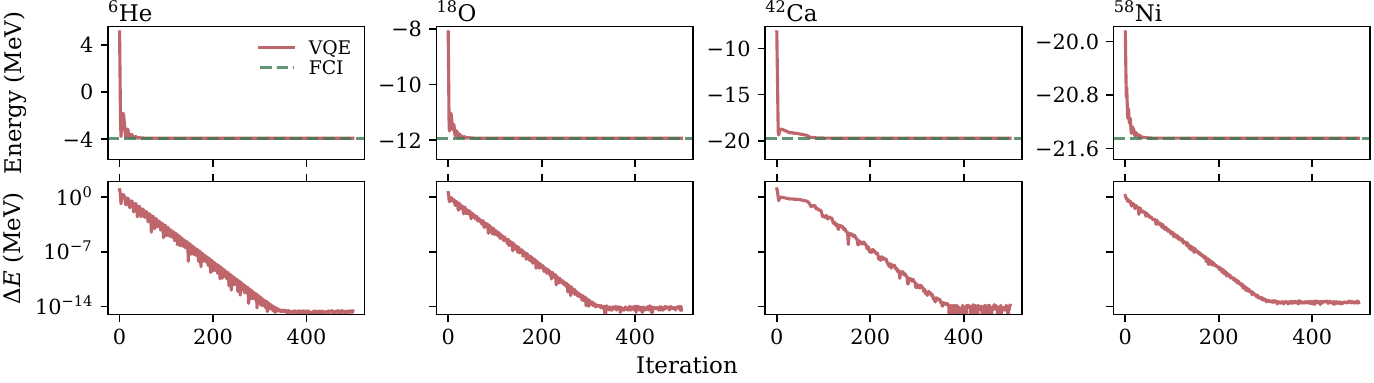}
\caption{\label{fig:VQE_pUCCD}
VQE results for representative two-valence-neutron systems using the pUCCD
ansatz. The systems ${}^{6}$He, ${}^{18}$O, ${}^{42}$Ca, and ${}^{58}$Ni are
described with CKpot, USDB, GXPF1A, and JUN45 interactions, respectively.
The variational parameters correspond to pair-excitation amplitudes implemented
with Givens-rotation gates. The upper panels compare VQE energies with
the shell-model reference (dashed). The lower panels show the energy
difference from DOCI on a logarithmic scale.
}
\end{figure*}

This appendix illustrates NuQuLib's variational interface using
mapped nuclear Hamiltonians, parameterized circuits, and a classical optimizer.
Figure~\ref{fig:VQE_pUCCD} shows results for two-valence-neutron systems
with the pair Unitary Coupled Cluster Doubles (pUCCD) ansatz.
These small examples demonstrate the software interface; they do not
establish the performance or scalability of VQE for more complex systems.

The pUCCD ansatz is constructed from pair excitations of time-reversal pairs
of single-particle orbits, and is therefore a natural choice for even-even nuclei.
In the present implementation, the variational parameters are the amplitudes of
pair excitations, which are implemented with Givens rotation gates.
The parameters are classically optimized using the Adam optimizer~\cite{Adam_2014}.
We follow the same methodology as previous work~\cite{SYPRR2026}.

We consider ${}^{6}$He, ${}^{18}$O, ${}^{42}$Ca, and ${}^{58}$Ni as representative
two-valence-neutron systems built on top of inert ${}^{4}$He, ${}^{16}$O,
${}^{40}$Ca, and ${}^{56}$Ni cores, respectively. The adopted valence-space
interactions are CKpot~\cite{CohenKurath}, USDB~\cite{USDB},
GXPF1A~\cite{Honma2005}, and JUN45~\cite{JUN45}, respectively.
These examples provide a compact test bed for the variational interface because
the pUCCD ansatz can be equivalent to the full configuration interaction (FCI) solution for these valence two-neutron systems.

\bibliography{references}

\end{document}